\documentclass[dvips,traditabstract]{aa}

\usepackage{graphicx,amssymb,amsmath,hyperref,multirow}
\usepackage[usenames,dvipsnames]{xcolor}

\usepackage{rotating}
\usepackage{txfonts}
\usepackage{lscape}
\usepackage{amsmath}
\usepackage{natbib,twoopt}
\usepackage{longtable}
\bibpunct{(}{)}{;}{a}{}{,} 
\newcommandtwoopt{\citeyearads}[3][][]%
{\href{http://adsabs.harvard.edu/abs/#3}{\citeyear[#1][#2]{#3}}}
\hypersetup{colorlinks,linkcolor=red,citecolor=blue,urlcolor=blue}

\newcommand{\tab}[1]{Table~\ref{#1}}

\newcommand{\apo}{APOKASC}

\begin{document}

\title{{Cannibals in the thick disk: \\the young $\alpha-$rich stars as evolved blue stragglers
}
}
\author{
P.~Jofr\'e\inst{\ref{ioa},\ref{port}}
\and A.~Jorissen\inst{\ref{ulb}}
\and S.~Van~Eck\inst{\ref{ulb}}
\and R.~G.~Izzard\inst{\ref{ioa}}
\and T.~Masseron\inst{\ref{ioa}}
\and K.~Hawkins\inst{\ref{ioa}}
 \and G.~Gilmore\inst{\ref{ioa}}
 \and C.~Paladini\inst{\ref{ulb}}
 \and A.~Escorza\inst{\ref{ulb},\ref{KU}}
 \and S.~Blanco-Cuaresma\inst{\ref{genf}}
 \and R.~Manick\inst{\ref{KU}}
	}

\authorrunning{Jofr\'e et al. }
\titlerunning{Evolved blue stragglers in the thick disk}
\offprints{ \\ 
P. Jofr\'e, \email{pjofre@ast.cam.ac.uk}}

\institute{
Institute of Astronomy, University of Cambridge, Madingley Road, CB3 0HA, Cambridge, United Kingdom\label{ioa}
\and 
	N\'ucleo de Astronom\'ia, Facultad de Ingenier\'ia, Universidad Diego Portales, 
         Av. Ej\'ercito 441, Santiago, Chile \label{port}	
\and
Institut d'Astronomie et d'Astrophysique, Universit\'e Libre de Bruxelles, Campus Plaine C.P. 226, Boulevard du Triomphe,\\ B-1050 Bruxelles, Belgium\label{ulb}
\and
Institute of Astronomy, KU Leuven, Celestijnenlaan 200D, B-3001 Leuven, Belgium\label{KU}
\and
	 Observatoire de Gen\`eve, Universit\'e de Gen\`eve, CH-1290 Versoix, Switzerland \label{genf}  
	}

   \date{}

\abstract{Spectro-seismic measurements of red giants enabled the recent discovery of stars in the thick disk that are more massive than $1.4~\mathrm{M}_\odot$. 
While it has been claimed that most of these stars are younger than the rest of the typical thick disk stars, we show evidence that they might be  products of mass transfer in binary evolution, notably evolved blue stragglers.   We took new measurements of the radial velocities in a sample of 26 stars from APOKASC, 
 including 13  ``young" stars 
and 13 ``old" stars with similar stellar parameters but  with masses below $1.2~ \mathrm{M}_\odot$ and found that more of the ``young" stars { appear to be} in binary systems with respect to the ``old" stars. { Furthermore, we show that the ``young" stars do not follow the expected trend of [C/H] ratios versus mass for individual stars. However, with a population synthesis of low-mass stars including binary evolution and mass transfer, we can reproduce the observed [C/N] ratios versus mass.}  Our study shows how asteroseismology of solar-type red giants  provides us with a unique opportunity to study the evolution of field blue stragglers after they have left the main-sequence. 
}
\keywords{Stars }
\maketitle
 
\section{Introduction}

Sixty years ago, a population of stars bluer than the turn-off was found in the globular cluster M3 \citep{1953AJ.....58...61S}. These stars appear to be younger than the bulk of the rest of the cluster stars.  Nowadays it is clear that blue straggler stars (BSSs) are the product of  coalescence or mass exchange in binary stars \citep{2008MNRAS.384.1263C, 2008MNRAS.387.1416C, 2015ebss.book.....B}.\\
\indent Most  studies on BSSs focus on stellar clusters because they  comprise a population of stars of the same age, chemical composition and distance, making the identification of the turn-off straightforward. On the other hand,  in the Galactic field the identification of BSSs is almost impossible, except perhaps the Galactic inner halo, which has a dominant population that is co-eval.  Thus, the halo has a well-defined turn-off colour  \citep{1996MNRAS.278..727U, 2001AJ....122.3419C, 2011A&A...533A..59J},  facilitating the identification of blue main-sequence stars. Based on the fraction of BSSs of globular clusters,  \cite{1996MNRAS.278..727U}  claimed that these blue metal-poor stars must be younger with an extragalactic origin. Radial velocity (RV) monitoring of these stars over several years showed that they are almost all in binary systems  \citep{2000AJ....120.1014P, 2001AJ....122.3419C}. This  implied that the majority of these stars were not young but ordinary BSSs, which became more massive because they gained mass through a mass transfer across the binary system.  \\
\indent The much higher frequency of BSSs in the Galactic field with respect to stellar clusters has been recently confirmed \citep{2015ApJ...801..116S}. This is attributed to the greater rate of wide binary disruption in high density environments such as globular clusters \citep{2004ApJ...604L.109P}. Hence,  blue stragglers are more likely to exist in the Galactic field, which is exactly  where they are most difficult to find. \\
%

\begin{table*}[!t]
\caption{Pairs of young  ($Y$) and old ($O$) stars are grouped together. Their 2MASS name and $K_s$ photometry, as well as their stellar parameters, masses and [C/N] abundances are listed. Columns $RV$ and $\sigma RV$ list the mean RV and the corresponding standard deviation. The $\chi^2$ of the RV measurements (adopting an uncertainty of 0.22~km~s$^{-1}$ for all measurements) and the probability $Prob$ of the star being in a binary system (in fact, $Prob$ is the probability integral of the $\chi^2$ distribution from 0 to the observed $\chi^2$ value for $Nvis - 1$ degrees of freedom) are tabulated as well. $Nvis$ denotes  the total number of visits (including APOGEE and HERMES) for each star and $tspan$ is the number of days between the first and the last observation.  }\label{28stars}
\small
\hspace{-0.6cm}
\begin{tabular}{|c|c|ccccccccccccc|}
\hline
Star & 2MASS name & $K_s$ & $T$eff & $\log g$ & [M/H] & Mass &[C/N]& $\sigma$[C/N]& $\bar{RV}$& $\sigma RV$& $\chi^2$& $Nvis$   & $Prob$ & $tspan$\\
 & &mag & K & cgs & dex & M$_\odot$ & dex & dex & km/s & km/s & & & & days\\
\hline 
\hline
$Y1$ &  J19083615+4641212 & 9.84 &4739 &2.71&-0.40 &1.67  &  0.10 &    0.08 &           -5.80 &       0.08 &       1.18 &  9 &  0.00 & 1718\\
$O1$ &  J19031115+4753540& 9.78  &4606&2.53&-0.45 & 0.95&      -0.13 &    0.11 &         5.29 &       2.43 &    608.09 &  6 &  1.00 & 1756\\
\hline 
$Y2$ & J19101154+3914584& 9.31&4741&2.50&-0.36& 1.56 &       -0.23 &    0.28 &           6.25 &       0.21 &        4.80 &  6 &  0.56 &1757\\
$O2$ & J19055465+3735053 & 9.26 &4756&3.40&-0.43&1.02&       -0.33 &    0.19 &        -16.38 &       0.04 &        0.20 &  8 &  0.00 &1759\\
\hline 
$Y3$ & J19081716+3924583 & 9.31 &4744 & 3.06  & -0.23 & 1.57&       0.03 &    0.11 &    -83.54 &      0.21 &       5.66 &   7 &  0.54 &1753\\
$O3$ &  J19313671+4532500 & 9.68 &4724&3.28&-0.18& 0.92&            0.21 &    0.03 &    -26.97 &      3.36 &  1165.68 &  6 &  1.00 &1577\\
\hline 
$Y4$ &  J19093999+4913392 & 9.24 &4785&2.50&-0.52& 1.38&    -0.40 &    0.28 &          -73.23 &       3.38 &     473.10 &  3 &  1.00 & 1692\\
$O4$ &  J19370398+3954132 & 11.44 &4774&2.381&-0.50& 1.04&       -0.09 &    0.10 &  -151.26 &       0.12 &         0.89 &  4 &  0.17& 1408\\
\hline 
$Y5$ &  J19032243+4547495 & 9.89 &4729&2.30&-0.19& 2.02&      -0.83 &    0.43 &         -86.29 &      0.11 &        1.60 &  8 &  0.02& 1539\\
$O5$ &  J19023483+4444086 & 8.48 &4695&2.43&-0.18& 1.00&     0.03 &    0.09 &              5.98 &      0.08 &        0.84 &  7 &  0.01& 1538\\
\hline 
$Y6$ &  J19455292+5002304 & 8.83  &4812&2.53&-0.51& 1.54&      -0.64 &    0.41 &        -22.64 &       4.88 &  2955.25 &  7 &  1.00& 1529\\
$O6$ & J19151828+4736422 & 10.01 &4819&2.38&-0.50& 0.90&      -0.28 &    0.23 &        -30.93 &       0.08 &        0.62 &  6 &  0.01& 1724\\
\hline 
$Y7$ &J18553092+4042447 & 10.74 &4849&2.51&-0.43& 1.63 &      -0.43 &    0.29 &     -38.40 &       1.71 &        300.70  &  6 & 1.00& 1724\\
$O7$ &J19411280+5009279 &  9.72 &4792&2.39&-0.43& 0.92 &      -0.34  &    0.22 &        1.97 &       0.04 &           0.21  &  6 &  0.00& 1529\\
\hline 
$Y8$ &  J19102133+4743193 & 8.55 &4676&2.4&-0.30& 1.39&     -0.25 &    0.24 &        -43.17 &       1.80 &        334.06 &  6 &   1.00& 1721\\ 
$O8$ &  J19084355+4317586 & 9.73 &4669&2.39&-0.30& 1.03&    -0.16 &    0.08 &       -46.69 &       0.24 &            7.32  & 7 &   0.71&1572\\
\hline 
$Y9$ &J18540578+4520474 &  8.68 &4229&1.55&-0.29& 1.81&       -0.60 &    0.36 &      -35.46 &       3.84 &    2443.43   &  9 &  1.00&1538\\
$O9$ &  J19254985+3701028 & 8.70 &4312&1.64&-0.23& 0.93&     -0.29 &    0.21 &       -69.99 &       0.37 &         5.29   &  8 &  0.37&1789\\
\hline 
$Y10$ &J19093801+4635253 & 8.89  &4435&1.85&-0.20& 1.49&       -0.53 &    0.30 &    -57.03 &       0.10 &       0.81 &  5 &  0.06 & 1721\\
$O10$ & J19382435+3847454 & 8.87 &4411&1.78&-0.07& 1.01&      -0.51 &    0.25 &        6.63 &       0.34 &      11.80&  6 &  0.96 & 1784\\
\hline 
$Y11$ &  J19052620+4921373 & 9.88 &4669&2.46&-0.17& 1.43&     -0.37 &    0.28 &      -26.80 &       0.08 &         0.50  &  5 &  0.03 & 1754\\
$O11$ &J19551232+4817344 &  8.21 &4644&2.42&-0.07& 1.02&       -0.27 &    0.14 &      -6.10 &        0.09 &         0.83  &  6 &  0.03& 1531\\
\hline 
$Y12$ &  J19374569+3835356 & 8.41 &4572&2.57&-0.08& 1.49&     -0.15 &    0.20 &      -47.64 &       0.51 &      37.87 &  8 &  1.00&1778\\
$O12$ &  J19070280+4530112 & 9.56 &4446&1.87&-0.05& 0.94&       -0.43 &    0.23 &     28.99 &       0.14 &        3.10 &  9 &  0.07 & 1784\\
\hline 
$Y13$ &J19024305+3854594 & 6.70 &4601&2.47&-0.02& 1.42&     -0.04 &    0.11 &     -62.27 &       0.14 &        2.45 &  7 &  0.13&1754\\ 
$O13$ & J19513344+4617498 & 8.86 &4602&2.38&-0.02& 0.96&       0.07 &    0.03 &   -41.43 &       0.09 &        0.96 &  7 &  0.01&1532\\
\hline 
\end{tabular}
\end{table*}%

\indent A new era of hunting mass-transfer products has arrived thanks to asteroseismic campaigns like {\it Kepler} \citep{2010Sci...327..977B} and {\it CoRoT} \citep{2006cosp...36.3749B} that allow us to measure masses directly from stellar oscillations \citep{2011Sci...332..213C}. Chemical abundances obtained via spectroscopy from e.g. the APOGEE survey  \citep{2015AJ....150..148H} allow us to assign a Galactic population to the stars.  The thick disk population, in particular, has low iron ($\mathrm{[Fe/H]}\sim -0.5$), but high $\alpha$ content ($\mathrm{[}\alpha\mathrm{/Fe]} > 0.2$). Furthermore, it is now well established that the thick disk is an old population \citep{2013A&A...560A.109H, 2015MNRAS.453.1855M}, and thus  no young stars are expected to belong to this population. Interestingly,  asterosismic data revealed that some of the thick disk giant stars were relatively massive ($M>1.4~\mathrm{M}_\odot$).   \cite{martig15} and  \cite{2015A&A...576L..12C} concluded that those must be young, revealing a paradigm nowadays referred to as the ``young $\alpha$-rich stars". This paradigm might be another case in which  BSSs are being misinterpreted  for young stars. \\
\indent The BSS formation scenario for these stars is rejected by \cite{martig15} partly based on the rather low specific frequency of BSS in  clusters. However, as mentioned above, this frequency is likely higher in the field. Their further argument is that their stars  are selected to be constant in RV, namely the stars were taken from \cite{2014A&A...564A.115A}, who selected only stars for which the scatter in the radial velocities of multiple visits is below $1$~km~s$^{-1}$. 
However,  6  out of the 14 young $\alpha-$rich stars had only one RV measurement. In the remaining stars, the RV scatter is obtained from a sample of observations that span at most one month  (see the observation dates of each target in the Obs. date column on the left side of Table~\ref{RV_individual}). Because binary BSSs  { have orbital periods which cluster around $10^3$~d, with just a few having periods shorter than 10~d} \citep[e.g.][]{2000AJ....120.1014P, 2008MNRAS.387.1416C, 2012AJ....144...54G},  the time span of the measurements considered in \cite{martig15} is too short to demonstrate that these peculiar stars have no companion.  For this reason, in 2015 and 2016 we obtained new measurements of their RVs, which come 3-5 years after the last APOGEE visit,  with the aim of testing possible long-period binaries and  the BSS scenario.

\section{Data}\label{rvs}

We selected 13 of the ``young" stars ($Y$) of \cite{martig15}. They are targets of the \apo\ catalogue \citep{2014ApJS..215...19P}, which is a joint project between APOGEE and {\it Kepler},  and have  masses above 1.4~$\mathrm{M}_{\odot}$. The spectra belong to the 12th data release (DR12) of APOGEE \citep{2011AJ....142...72E}. The RVs determination is described in \cite{2015AJ....150..173N} and is based on cross correlation. We considered the masses derived using the scaling relations listed in Table~5 of \cite{2014ApJS..215...19P}. \\
\indent In addition, we included a comparative sample of ``old'' stars ($O$). For each of the $Y$ stars we looked for another star in the \apo\ sample with similar atmospheric stellar parameters, but with a mass below $1.2~  \mathrm{M}_\odot$.  { We are aware that this mass limit might still be too large for an expected age of 8~Gyr or more for typical thick-disk stars, but the mass distribution of the $\alpha-$rich metal-poor stars of the \apo\ sample reaches that limit \citep[see e.g.][]{martig15}.  Here our $O$ stars are meant to be clearly part of the bulk of the ``normal'' population, and therefore we selected stars with masses  such that they fall below $1.2~  \mathrm{M}_\odot$ when considering the errors. }  Our stellar parameters and chemical abundances are derived in \cite{2016arXiv160408800H}. Briefly,  temperature and gravity were fixed to derive metallicities and chemical abundances using the BACCHUS code, very similar to \cite{2014A&A...564A.133J, 2015A&A...582A..81J}. The differences here are that the code and the line list were designed to analyse the infrared spectra of APOGEE, and the stellar parameters were based on the seismic and photometric information.  
Our sample consists of 26 stars,  13 $Y$ and 13 $O$, whose main properties are listed in Table~\ref{28stars}. 

\subsection{New RV measurements with HERMES}

The new RV data are obtained with the HERMES spectrograph mounted on the 1.2m Mercator telescope,  at the Roque de Los Muchachos Observatory,  La Palma  \citep{2011A&A...526A..69R}. Two main observing runs were carried out, one in July-August 2015 and another one between May and July 2016.  The spectrograph covers the optical wavelength range from $380$ to $900~\mathrm{nm}$ with a spectral resolution of about 85,000. RVs were determined by cross-correlation with a mask of Arcturus covering the range $\sim480 - 650~\mathrm{nm}$. This avoids both telluric lines at the red end, as well as the crowded and poorly exposed blue end of the spectrum. \\
\indent The exposure times were calculated according to the brightness of the star in order to achieve a SNR of 15. This is normally sufficient to obtain a well-defined cross-correlation function that yields an internal precision of a few $\mathrm{m/s}$ on the RV \citep{2016A&A...586A.158J}. 
The RVs from both APOGEE and HERMES are listed in Table~\ref{RV_individual}.

\subsection{Radial velocity uncertainties}\label{rv_error}

The errors listed in Table~\ref{RV_individual} are  $1\sigma$ uncertainties obtained from the cross-correlation function. For instrumental uncertainties, the APOGEE team reports an error distribution peaking at $100~\mathrm{m/s}$ estimated from the typical RV scatter of stars with multiple visits \citep{2015AJ....150..173N}. 
For HERMES it has been shown that typical uncertainties are of $55~\mathrm{m/s}$, and in few cases uncertainties can be up to $100~\mathrm{m/s}$. They are obtained from the monitoring of RV standards over several years \citep{2016A&A...586A.158J}. \\
\indent In this work, we are interested in setting a threshold on RV variations above which binary stars might be held responsible for the RV variations given the measurement uncertainties. We know that stellar radial velocities are affected by  jitter caused by pulsations and other inhomogeneities in their atmosphere \citep{1988A&A...198..187J, 2003A&A...397.1151S, 2009A&A...498..627F}. Although large effects happen at the tip of the red giant branch, their impact on less luminous red giants is poorly understood 
\citep{1988A&A...198..187J, 2008AJ....135..196C}. In addition, because we  assess the binary nature of the stars considering measurements from different instruments, we must  account for systematic offsets in the instrumental zero point. While the HERMES pipeline has been corrected to meet the \cite{1999ASPC..185..367U} list of standard velocity stars,  the APOGEE RV pipeline has reported an offset of $355\pm33~\mathrm{m/s}$ for DR12 with respect to stars in common with \cite{2012arXiv1207.6212C}, which differs from \cite{1999ASPC..185..367U} by an average offset of $63\pm72~\mathrm{m/s}$, { with some sensitivity to the stellar colour, encapsulated in the standard deviation}. Thus, for analysing both samples, we corrected the APOGEE observations by subtracting $355+63 =418~\mathrm{m/s}$ to the values indicated in \tab{RV_individual}. { The uncertainty on the zero-point correction amounts to 79.2~m/s, which is obtained from the root-mean-square of the standard deviations given above.  This uncertainty will hinder the detection of binaries with amplitudes of that order. 
The criterion for a star being in a binary system will strongly rely on the value adopted for the velocity uncertainty, but that value is difficult to fix a priori, because of the uncertain amount of velocity jitter in the considered stars. We will therefore fix the error uncertainty a posteriori, as explained in Sect.~\ref{velocities}. 
}


\begin{figure}

\includegraphics[scale=0.45]{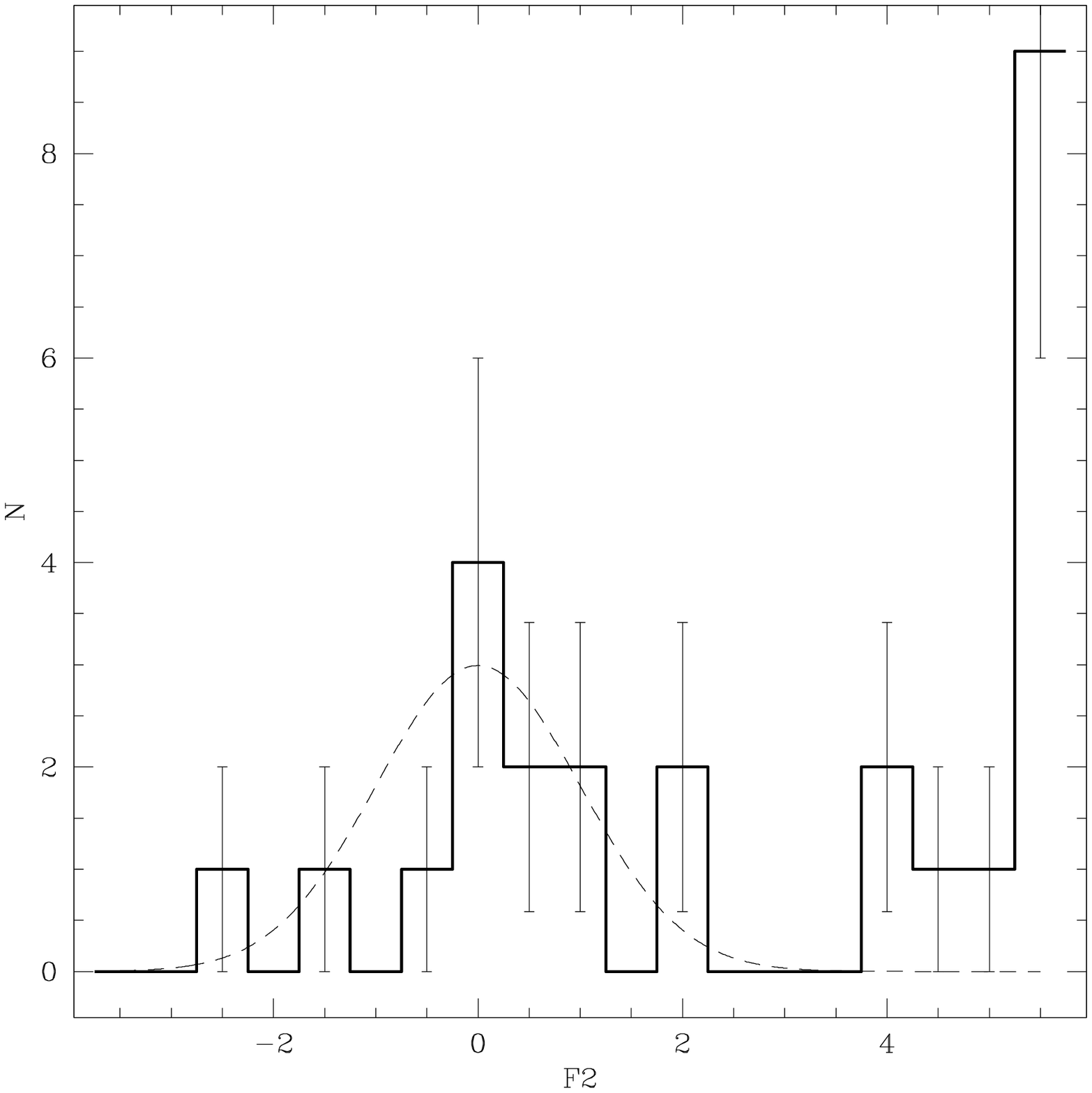}
\includegraphics[scale=0.45]{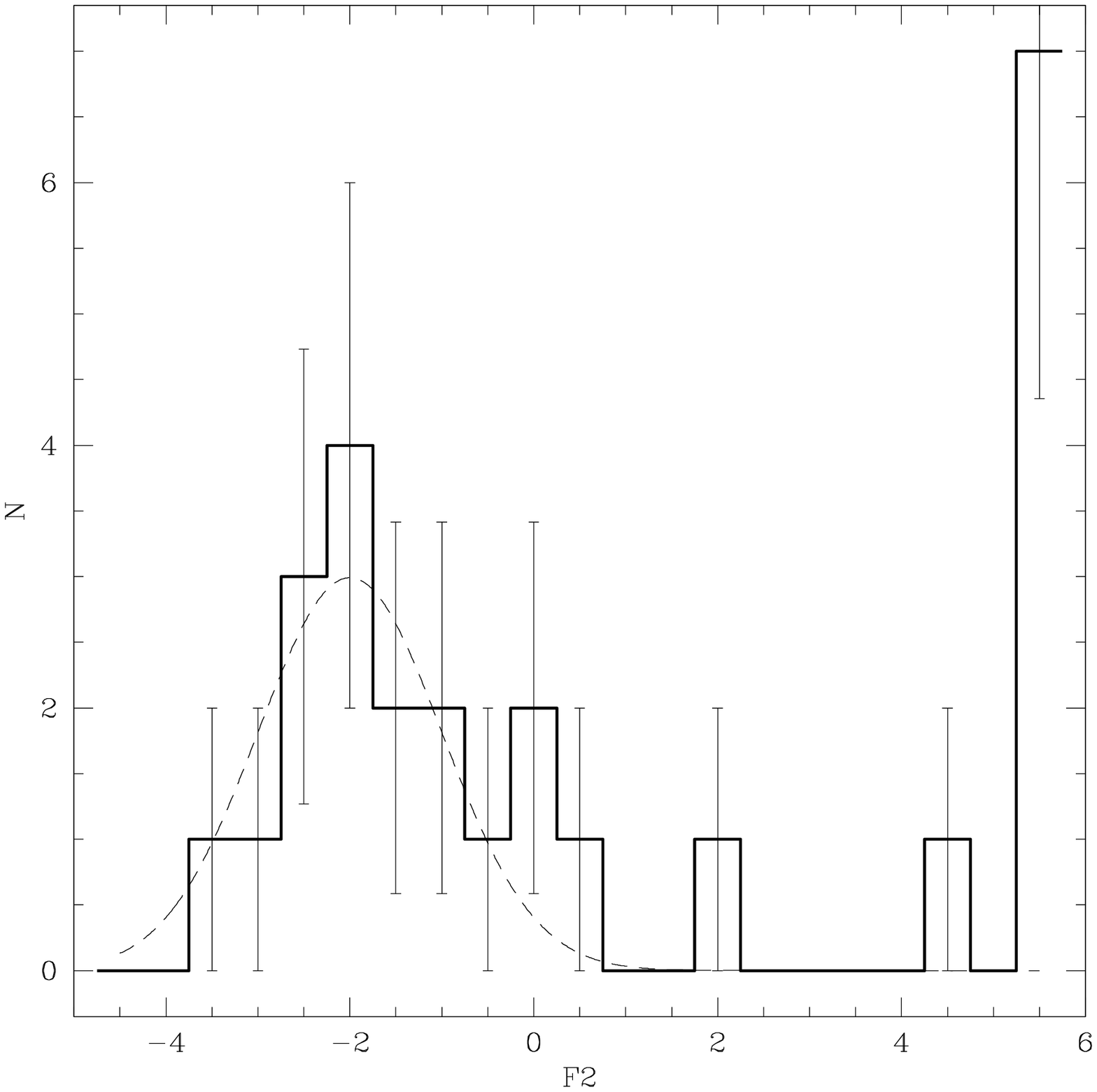}
\caption{ Top panel: The distribution of $F2$ values (see text and Eq.~\ref{Eq:f2}) for the combined sample of $O$ and $Y$ stars, by adopting an uncertainty of 88~m/s on every measurement. The dashed line is a reduced Gaussian distribution (mean of 0 and standard deviation of 1). Bottom panel: Same as top, adopting an uncertainty of 220~m/s on every measurement. The dashed line is a Gaussian distribution of mean -2  and unit standard deviation.}
\label{Fig:F2}
\end{figure}

\section{Results}

\subsection{Radial velocities and binary frequency}\label{velocities}


{ In the present section, we describe first how we derived an estimate of the uncertainty on our radial-velocity measurements, and then how a given star is flagged as being a binary. 

\subsubsection{RV uncertainty}

Assuming that the error on each measurement is a fixed value $\sigma$, we may 
calculate the $\chi^2$ value based on all $N$ observations for a given star,}
\begin{equation}
\label{Eq:chi2}
\chi^2 = \sum_{i=1}^N  \frac{(RV_i- \bar{RV})^2}{\sigma^2} \, ,
\end{equation}
\noindent where $RV_i$ is the RV measurement of the visit $i$, { corrected for the zero point of the APOGEE measurements by} $-0.418~\mathrm{km/s}$, 
and $\bar{RV}$ is the mean RV value computed from the set of  $N$ observations. 
{  If the RV uncertainty is correctly estimated, the $\chi^2$ values should follow a chi-square
distribution. However,  since each star has different number of available observations,  these distributions have different degrees of freedom, and therefore it is better to apply this concept on the standardised reduced chi-square $F2$, defined as \citep{Wilson1931}:
\begin{equation}\label{Eq:f2}
F2 = \left(\frac{9\nu}{2}\right)^{1/2}\left[\left(\frac{\chi^2}{\nu}\right)^{1/3}+\frac{2}{9\nu}-1\right],
\end{equation}
where $\nu$ is the number of degrees of freedom of the $\chi^2$ variable. The transformation of ($\chi^2, \nu$) to $F2$ eliminates the inconvenience of having the distribution depending on the additional variable $\nu$, which is not the same over the whole sample of our stars. $F2$ must follow a normal distribution with zero mean and unit standard deviation, provided  that $\sigma$ is correctly estimated. If  $\sigma$ is overestimated, then the obtained $\chi^2$  values become too small or the $F2$ distribution has a negative mean instead of null mean.

\begin{figure}[!t]
\hspace{-1.3cm}
\includegraphics[scale=0.50]{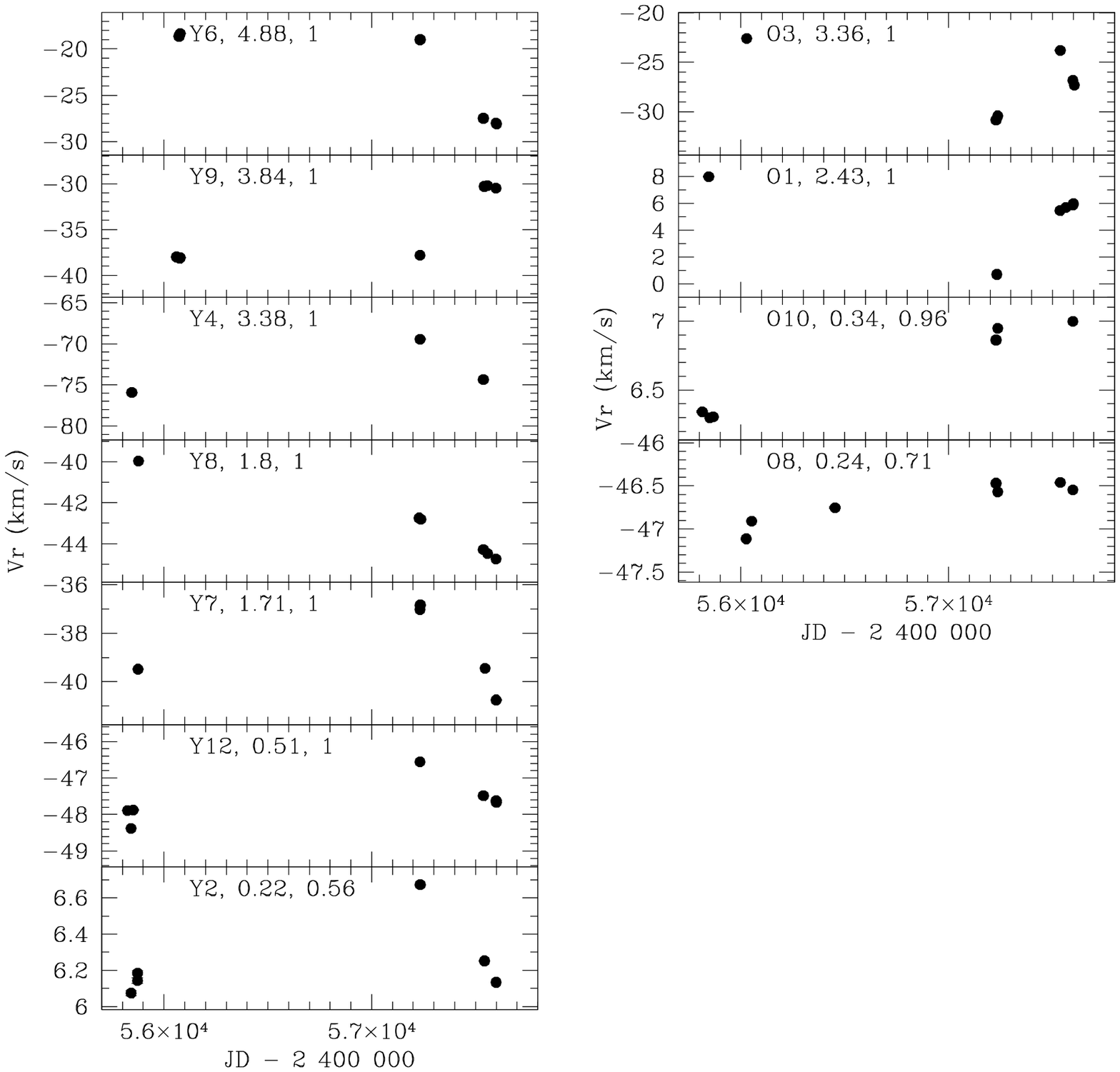}
\vspace{-0.5cm}
\caption{ Radial-velocity curves for the stars flagged as binaries (left column: Y stars; right column: O stars). The labels in each panel give the star number, $\sigma(Vr)$ value (in km/s), and probability of being a binary.}
\label{Fig:binaries}
\end{figure}

 The value $\sigma = 90$~m/s appears to yield a  $F2$ distribution adequately represented by a normal distribution with zero mean and unit standard deviation (see upper panel of Fig.~\ref{Fig:F2}). This value agrees well within the uncertainties of the HERMES and APOGEE measurements (Sect.~\ref{rv_error}). It also means that no jitter is expected to be present, since the measurement uncertainty suffices to adequately match the distribution of errors.  The tail of large $F2$ values in Fig.~\ref{Fig:F2} cannot be explained by the error distribution,  or by an inadequate zero-point offset between the APOGEE and HERMES data. These stars are our best binary candidates. 
 
    We stressed in Sect.~\ref{rv_error} that there is an uncertainty of  79~m/s due to zero point offset, which is comparable to the measurement uncertainty explained above. To be conservative and to allow for a possible error on the zero-point offset, not all stars in the large-$F2$ tail should be flagged as definite binaries. Therefore, the above process has been repeated, but adopting this time a larger $\sigma$ of 220~m/s. Although  the normal distribution has a negative mean in this case (see bottom panel of Fig.~\ref{Fig:F2}) we found  that this value allows us to set aside with better confidence those stars for which velocity variations  may be due to either binarity or to an error on the zero-point offset (discussed in more detail below).

    \subsubsection{Binary probability}
    
         To be more quantitative,} the probability $Prob$ of the star being a binary was calculated from the $\chi^2$ distribution considering $N-1$ degrees of freedom and  $\sigma = 220$~m/s in Eq.~\ref{Eq:chi2} \citep[see also][for details]{2005ApJ...625..825L}.  These values, together with the averaged RV and its standard deviation, are listed in Table~\ref{28stars}. 
The full time span of the measurements is indicated in the last column of \tab{28stars}, showing that we deal with time spans of more than 1400~days for all stars. 

\begin{figure}[!t]
\hspace{-0.5cm}
\vspace{-4cm}
\includegraphics[scale=0.50]{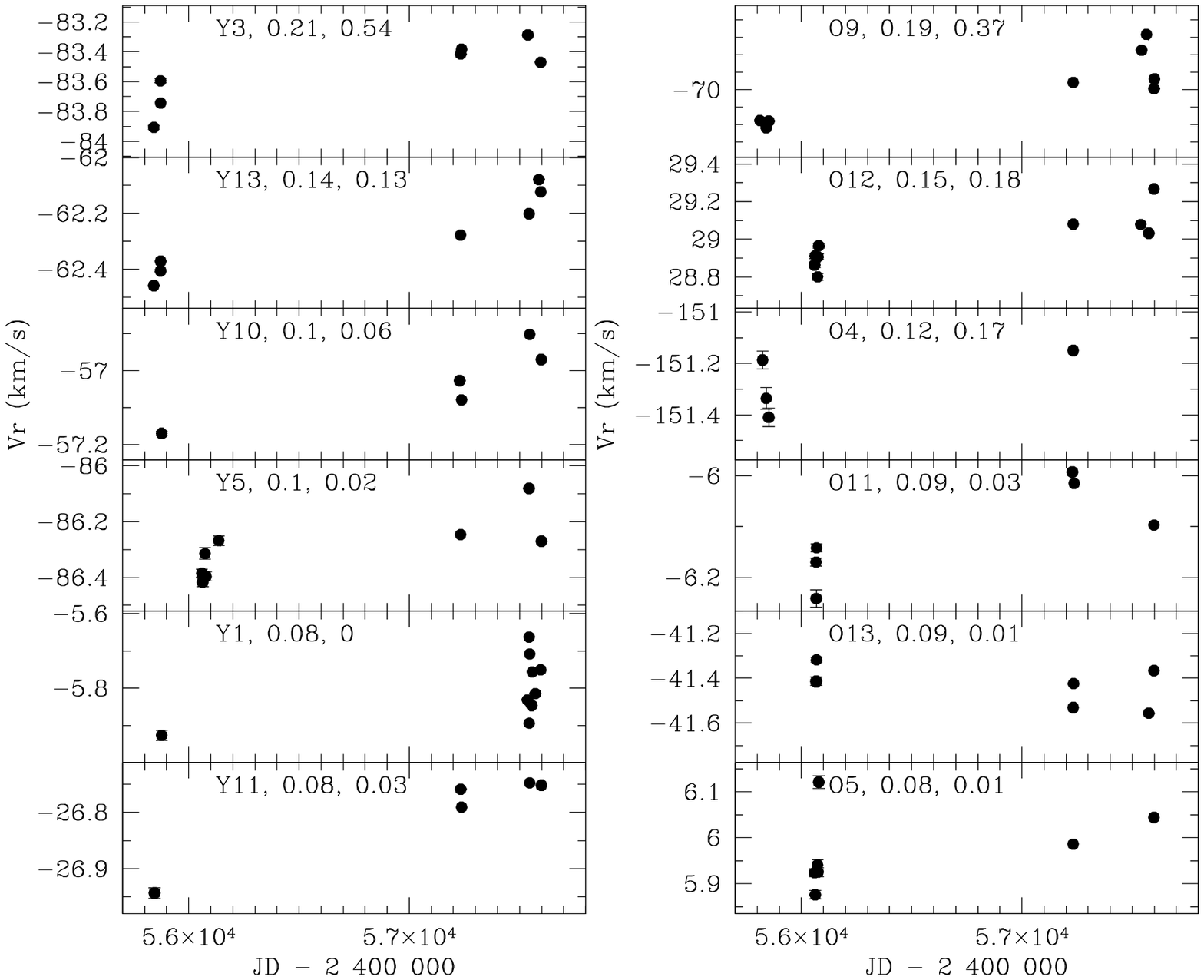}
\vspace{-2.5cm}
\hspace{-1.0cm}
\includegraphics[scale=0.50]{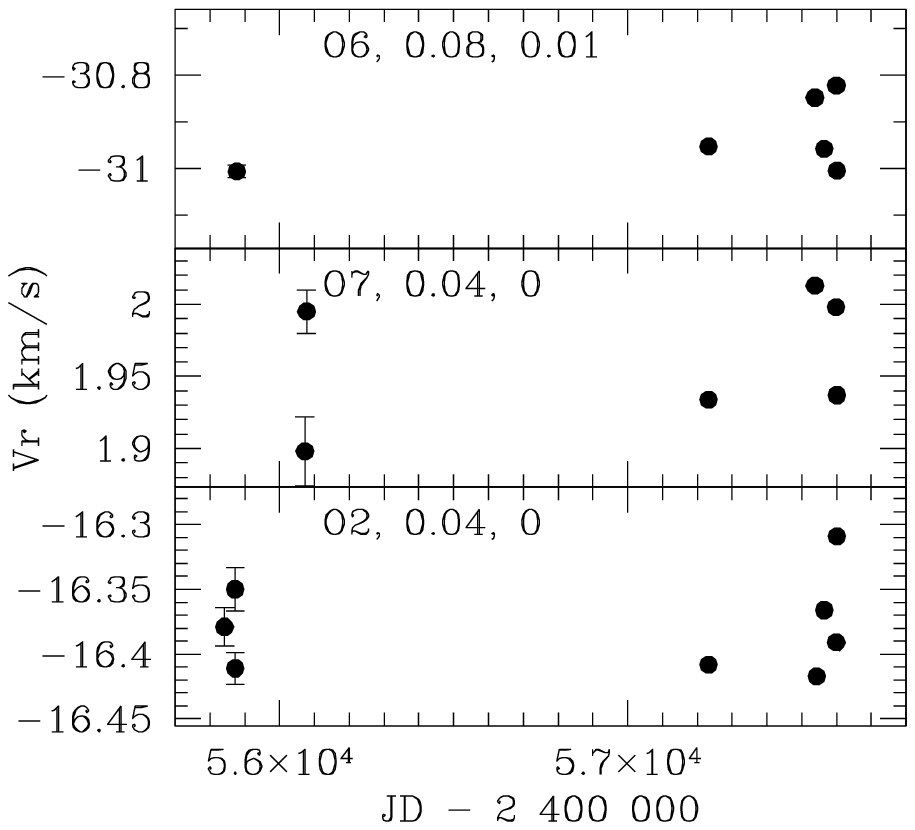}
\caption[]{ Same as Fig.~\ref{Fig:binaries} for stars with no firm evidence for orbital motion. For most of these stars, a change in the zero-point offset 
by only 0.1~km/s (as compared to the value $0.42\pm0.08$~km/s used here) would considerably alter the visual impression of the existence or not of a long-term trend between the APOGEE and HERMES data.}
\label{Fig:nonbin}
\end{figure}

\indent  The stars with 100\% confidence of being binaries ($Prob = 1$) correspond to  {  $O1$, $O3$, $Y4$, $Y6$, $Y7$, $Y8$, $O8$, $Y9$, $O10$, and  $Y12$. Their radial-velocity curves are shown in Fig.~\ref{Fig:binaries}. We may add $Y2$ in the figure: despite the fact that the confidence level of $Y2$ being a binary is only 56\% (just below the 1$\sigma$ confidence level), the HERMES measurements alone show a clear trend with an amplitude of 0.6~km/s, which is well above the HERMES uncertainty of 0.22~km/s.

\begin{figure*}[!t]
\vspace{-0.5cm}
\hspace{-1.0cm}
\includegraphics[scale=0.65]{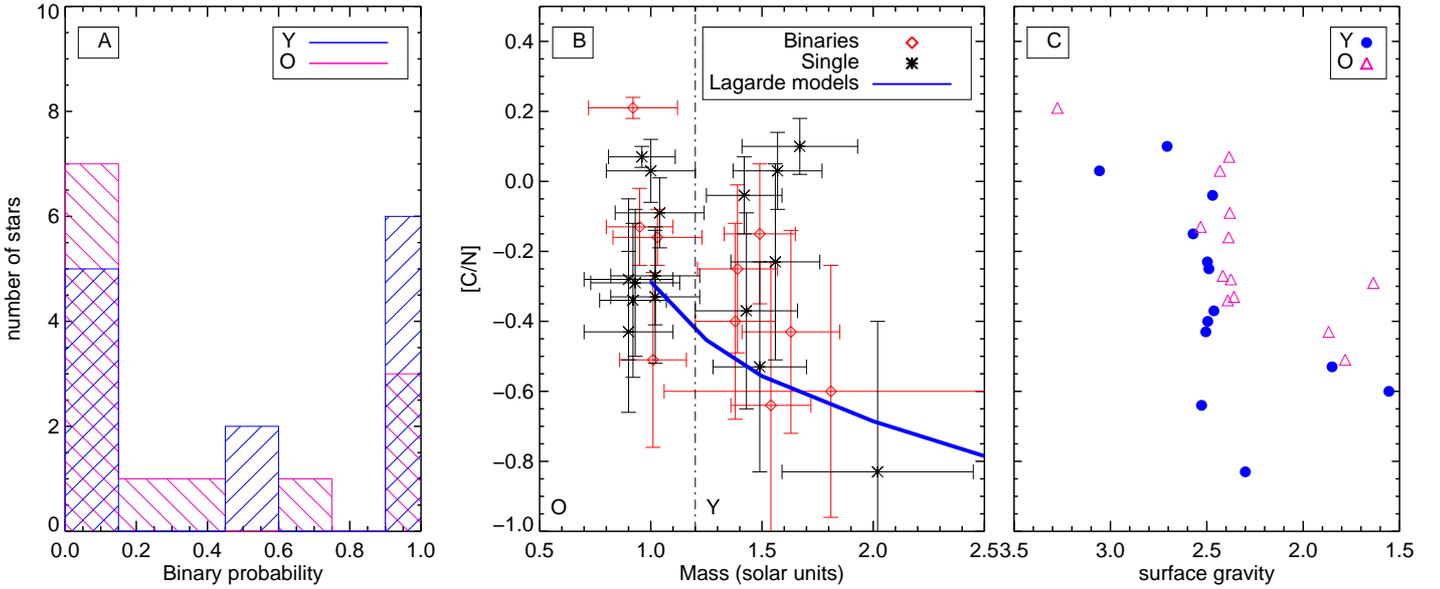}
\caption{Panel A: Distribution of stars as a function of binary probability. There are more $Y$ stars with high probabilities of being binaries than $O$ stars. Panel B: [C/N] abundances as a function of mass for our sample of stars. The vertical dashed line marks the mass threshold between $O$ and $Y$ stars. Red symbols denote stars with probabilities larger than 68\% of being binaries, while black symbols denote stars more likely to be single. The blue solid line indicates the theoretical relation between [$^{12}$C/$^{14}$N] and stellar mass in a post-FDU population, from \citet{2012A&A...543A.108L}.  There is no significant relation between [C/N] and binarity in the sample and several $Y$ stars do not follow the theoretical relation. Panel C: [C/N] abundances as a function of surface gravity. Blue circles correspond to the $Y$ stars and pink triangles represent $O$ stars. 
}
\label{cn_ratio}
\end{figure*}

The rest of the  stars show so far no evidence of being binaries. Their radial-velocity curves are shown in Fig.~\ref{Fig:nonbin}. This  conclusion is, however, somewhat 
dependent upon the adopted value of the zero-point offset. Some stars show  a long-term trend (e.g., $Y10$, $Y11$, and $Y13$) , but its amplitude is not much larger than the uncertainty on the zero-point offset, so that no firm conclusion can be drawn at this stage. More HERMES observations are planned for the future to reach a definite conclusion about the binary nature of the stars depicted on Fig.~\ref{Fig:nonbin}.  
In summary, 54\% of the sample of $Y$ stars are likely in binary systems, as compared to 31\% for  the $O$ stars.} In panel A of Fig.~\ref{cn_ratio}, we show the binary probability distribution of the $Y$ and $O$ samples, plotted with blue and pink colours, respectively.

 \indent Since this result is based on a small sample with few measurements, it is worth evaluating whether or  not  this difference of  binary frequencies between $Y$ and $O$ stars is statistically significant. To do so, we use the hypergeometric test as in \cite{2009A&A...498..479F}. We define $N_y = N_o = 13$ for the  number of $Y$ and $O$ stars, and $N_t = N_y + N_o = 26$ for the total number of stars.  { Among these, we have $x_y = 7$ $Y$ binaries, and $x_o = 4$ $O$ binaries. We further define $x_t = x_y + x_o = 11$. The frequency of binaries is then  $p_y = x_y/N_y = 0.54$, $p_o =  x_o/N_o = 0.31$ for the $Y$ and the $O$ samples, respectively, and $p_t = 0.42$ for the total sample.}
The expected number of binaries in the $Y$ sample ($\tilde{x}_y$) is computed from the total fraction of binaries applied to the number of $Y$ stars  ($\tilde{x}_y = p_t N_y = 5.46$). The corresponding standard deviation on that inference ($\sigma_{x_y}$) can be computed from the hypergeometric distribution given the total number of stars, the  number of binaries and the  number of $Y$ stars: $\sigma_{x_y} = [N_y p_t(1-p_t)(N_t - N_y)/(N_y-1)]^{1/2} = 1.85$. Finally, the significance of the difference between the expected and the observed binary frequencies $(c^2 = [(\tilde{x_y} - x_y)/\sigma_{x_y}]^2 = 0.69)$ may be computed from the $\chi^2$ distribution with one degree of freedom \citep[see][for details]{2009A&A...498..479F}, which in this case corresponds to { 40}\%.
Thus, given our data,  
there is a { 40}\% chance that  the observed difference
between the binary frequencies in the $O$ and $Y$ samples is simply due to statistical fluctuations. { Admittedly, the significance of the existence of a true difference between the $Y$ and $O$ populations is not high yet. }
In the remainder of this article, we will nevertheless 
assume that the binary frequencies are truly different in the $O$ and $Y$ samples, with the fraction of $Y$ binaries larger than the one of $O$ binaries.\footnote{ It is important to remark that it is also possible that the larger frequency of  RV variations among the $Y$ sample as compared to the $O$ sample could be due to differences in jitter related to mass. A definite answer on that question should await the demonstration of the Keplerian nature of the velocity variations for $Y$ stars, even though  the long-term trends already visible on Fig.~\ref{Fig:binaries} are clearly in favour of orbital variations.} { This conclusion is supported by the fact that the percentage of 54\%  binaries found in the $Y$ sample is especially large when compared to that found among comparison samples of K or M giants, where the percentage of spectroscopic binaries (observed under similar conditions as the $Y$ and $O$ samples discussed here) ranges between 6.3\% and 30\%, depending on the selection criterion of the sample \citep[see the discussion by][]{2009A&A...498..479F}. It may therefore be suspected that the large fraction of binaries (54\%) among the $Y$ sample is related to mass-transfer episodes responsible for the masses of $Y$ stars, whereas the binary frequency among the $O$ sample is similar to the upper range found among K giants.

}

\subsection{Carbon and Nitrogen}

{ 
As shown by \cite{2015MNRAS.453.1855M}, the [C/N] ratio in the APOGEE giants  can be used as an indication of their mass. The first dredge-up (FDU) mixes material from the CN-processed inner layers to the surface of the star, at an amount that is a function of the stellar main-sequence mass \citep{1965ApJ...142.1447I}.  Thus,  [C/N] anticorrelates with mass in a post-FDU population \citep{2012A&A...543A.108L,  2015A&A...573A..55T, 2015A&A...583A..87S, 2016MNRAS.456.3655M}. \\
\indent  We can check this correlation in the present sample, for which asteroseismic masses are available.
Panel B of Fig.~\ref{cn_ratio} shows our [C/N] ratio determinations as a function of stellar mass, indicating with red diamonds the stars with $Prob = 1$ (presumably binaries)  and with black asterisks the stars with $Prob < 1$ (presumably single stars).  The vertical dashed line represents the mass threshold used to define the $O$ sample, which have masses below $1.2~\mathrm{M}_{\odot}$ { (allowing for a 1$\sigma$ error bar)}. The blue solid line corresponds to the theoretical relation of \cite{2012A&A...543A.108L} for the post-FDU models with $Z=0.004$. To mimic the enhanced initial composition in [C/N] at low metallicity \citep{2015MNRAS.453.1855M}, we  added 0.1~dex to the model predictions. We plot this relation to qualitatively show the expected anti-correlation of [C/N] vs mass for single star, post-FDU models. 

Many binary stars in panel B of Fig.~\ref{cn_ratio} are clearly off this ([C/N], mass) anti-correlation,
which therefore does not seem to apply for young $\alpha$-rich stars (the $Y$ sample). 
But since young $\alpha$-rich stars are believed to be binaries, single star evolution models might not be relevant for them.

Moreover, as stated above, the [C/N] ratio should depend not only on mass, but also on the FDU occurrence. To distinguish pre-FDU stars from post-FDU stars, we plot in panel C of  Fig.~\ref{cn_ratio} the [C/N] ratio as a function of the surface gravity.
In the models of \cite{2012A&A...543A.108L}, the DUP tends to happen at $\log g$ around 3 \citep[see lower panel of Fig.~1 of][for different masses and metallicities]{2015MNRAS.453.1855M}, which is consistent with Panel C of Fig.~\ref{cn_ratio}, where the stars show a change in [C/N] around $\log g$ of 2.5. 

Therefore, if all stars with $\log g <$ 2.5 are indeed post-FDU objects, the fact that among those, $Y$ stars have lower [C/N] ratio than $O$ stars might, at first sight, seem puzzling in the framework of a mass transfer scenario.
Indeed, $Y$ stars are believed to have accreted [C/N] matter, in the past, from an $O$ star. Accounting for dilution, the [C/N] ratio of any $Y$ star could not be lower than the lowest [C/N] ratio observed for $O$ stars. 
In fact, population synthesis simulations (see next section) show that the mass transfer is likely to occur during the Hertzsprung gap, before the FDU, and so to pollute the companion with pristine gas, with [C/N]~$\sim 0$.
The decrease of the [C/N] ratio only occurs when the accretor itself experiences its FDU. 
Since the polluted star ($Y$) had, as a result of mass transfer,  a larger main-sequence mass, its FDU is expected to decrease
 more significantly its [C/N] than it does for stars from the $O$ sample.
It is therefore not surprising that post-FDU stars from the $Y$ sample tend to have lower [C/N] values than stars from the $O$ sample of similar gravity, as illustrated in panel C of  Fig.~\ref{cn_ratio}.

}

 \begin{figure}[!t]
\hspace{-1.0cm}
\includegraphics[scale=0.50]{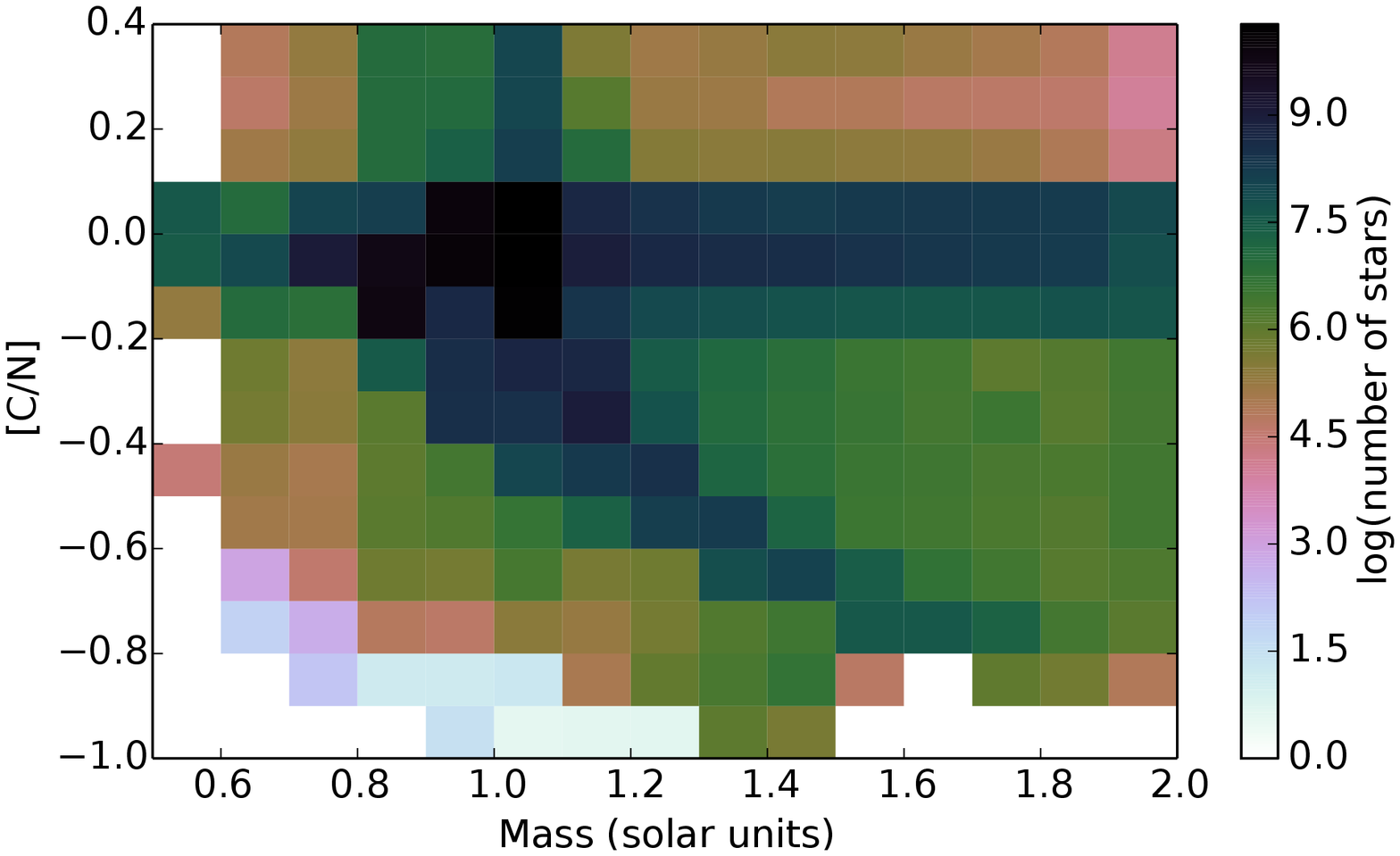}
\vspace{-0.5cm}
\caption{[C/N] vs mass in giant stars of our synthetic population by \citet[{\it in prep.}]{2004MNRAS.350..407I, 2006A&A...460..565I, 2009A&A...508.1359I} including initially 50\% of binary stars and mass transfer via RLOF and wind-RLOF.  The population has $Z = 0.004$ with all stars being older than 4~Gyr.} 
\label{cn_stats}
\end{figure}

\subsection{Theoretical predictions from a population synthesis model}\label{model}

Since the previous sections hint at binary systems playing a role in the occurrence of the $Y$ population,  we evaluated how binary interaction could account for increasing masses and altering binary orbits and surface abundances for the giant stars of the $Y$ sample.

The population synthesis model used here follows the description of \citet[{\it in prep}]{2004MNRAS.350..407I, 2006A&A...460..565I, 2009A&A...508.1359I}.  
  We simulate a stellar population with 50\% binary and 50\% single stars, with periods distributed according to \cite{1991A&A...248..485D} and a flat mass-ratio distribution. We comment here that this frequency of 50\% of binaries is only to include binaries in our sample and adopting another value does not change our conclusions.  { The main idea here is to show that if we do not include binaries, there will be no mechanism to obtain massive stars. The amount of resulting massive stars from the binary evolution depends on the fraction of initial binaries and single stars, as well as the period distributions of the binaries. This is discussed in depth in a complementary article on this subject (Izzard et al. {\it in prep.})}. Mass transfer in binaries follows the Roche Lobe overflow formulation  (RLOF) of \cite{2014A&A...563A..83C}, which depends on the mass ratio of the binary system $q$. Wind mass transfer is based on the wind-RLOF formalism of \cite{2013A&A...552A..26A, 2015A&A...581A..62A}.  Our synthetic population has a metallicity of $Z = 0.004$, which is consistent with the thick disk population. Because our goal is to test whether mass transfer can produce  $1.4~\mathrm{M}_\odot$ stars in the thick disk, our population contains only stars older than 4~Gyr, corresponding to single stars with masses below $1.2~\mathrm{M}_{\odot}$.   A complementary study to this article discusses in depth the different input physics of this model and how the model results compare to the observations of the \apo\ sample (Izzard et al, {\it in prep}). Here we only focus on a qualitative comparison,  showing that interacting binaries can also explain the characteristics seen in the stars of the $Y$ sample. 


In Fig.~\ref{cn_stats} we plot [C/N]  as a function of mass in our synthetic population using the colour scheme of \cite{2011BASI...39..289G}. The shading indicates the logarithm of the number of stars in a given [C/N]-mass bin. The first interesting aspect to notice is that, while we did not allow for  single stars with $\mathrm{M}>1.2~\mathrm{M}_\odot$ in the initial sample,  binary interaction creates stars with masses above this value.  Furthermore, the model reproduces
the standard { anticorrelation between [C/N] and mass}  
\citep[similar to that for single stars from ][shown in Fig.~\ref{cn_ratio}]{2012A&A...543A.108L}. It predicts 
 a secondary sequence at constant [C/N]. As discussed above, the latter is not predicted
in a single stellar evolution scenario. 

Some model stars have masses $\sim 1~\mathrm{M}_{\odot}$ and high [C/N]  in Fig.~\ref{cn_stats}. This is the locus 
occupied by post-mass-transfer giants in binary systems like CH,
barium and extrinsic S stars \citep{1942ApJ....96..101K, 1951ApJ...114..473B, 1990ApJ...352..709M, 1998A&A...332..877J,2004MNRAS.350..407I, 2009A&A...508.1359I}. They are products of wind mass transfer from an AGB donor which was able to synthesise
C and s-process elements. Since those stars are not present in our observed sample, we will
not discuss them further here (but it is definitely worth searching for them in the 
\apo\ sample).

\subsection{Discussion}

Stars with $M > 1.4~ \mathrm M_\odot$ found in the thick disk by {\it Kepler} and {\it CoRoT} have been interpreted as young stars. This challenges Galactic evolution models  because it is not easy to explain the presence of such young, $\alpha-$enhanced stars in the solar neighbourhood \citep{2013A&A...558A...9M}. One explanation is that they were formed in the region near the co-rotating bar, which is  where the gas can be kept inert for long time and expelled to outer regions by radial migration \citep{2014ApJ...781L..20M, 2015A&A...576L..12C}.  It is still not clear how to make the young $\alpha-$rich  stars migrate from the bar to the solar vicinity in the  lifetime of such young stars.  It is worth to comment that \cite{2014A&A...569A..17C} and \cite{2015A&A...580A..85M}
discuss the nature of the young $\alpha-$rich open cluster NGC 6705 { (of age $0.30\pm0.05$~Gyr and with [Mg/Fe]$ \sim $[Si/Fe]$ \sim +0.3$~dex for [Fe/H]~$= +0.10\pm0.06$~dex, based on the analysis of 21 stars)}. This cluster is more metal-rich than the stars analysed here and these authors propose that the $\alpha-$enrichment of this young cluster might be due to local enrichment of the molecular cloud from supernova explosions, instead of radial migration.

In this article we discuss whether binary stars could account for 
the (sub)giant stars dubbed ``young $\alpha-$rich". We
propose that they were blue straggler main-sequence stars or other binary stars that underwent mass-transfer in the past. 
This agrees with the recent suggestion of \cite{2016arXiv160101412B} based on a work on open clusters with eclipsing binary systems and asteroseismic observations.
This further agrees with the recent study of \cite{2015ApJ...807...82T}, who analysed anomalous rotational velocities in \apo\  and explained them as mass transfer products.  

The $Y$ stars could have been formed following a mass-transfer scenario,
as for other BSSs found in the field \citep{1964MNRAS.128..147M, 2015ebss.book...65P}.   Since BSSs  { have orbital periods which cluster around $10^3$~d} \citep{2012AJ....144...54G}, a time span longer than the few days spanned
by the data used by \cite{martig15} is required 
to assess their binary nature. Indeed, with observations spanning more than 1000 days, we already find more than 50\% of binaries among the $Y$ sample. 
More should be expected with an even longer time span.

Our results support the following formation scenario for
the stars belonging to the $Y$ sample: 
when the less massive component of the system is still on
the main sequence, it accretes matter from its more massive
companion donor star, as the latter evolves through the Hertzsprung gap. Since this phase
corresponds to a rapid increase of the stellar radius, for close-enough systems,  RLOF 
will occur.  
In this phase, the system may look like an Algol (once the mass ratio has been reversed), and 
a substantial amount of mass may be transferred from the donor to the accretor. 
Because the Hertzsprung gap happens before FDU, the
donor transfers pristine gas, i.e. with [C/N]$\sim 0$. Thus, as the
accretor gains mass, it moves horizontally to the right in Fig.~\ref{cn_stats}. At the same time, in the Hertzsprung-Russell diagram, the accretor becomes
a blue straggler.
Long after mass transfer has ceased, 
the accretor itself leaves the main sequence
to become a red giant. Before FDU, the giant preserves
its pristine [C/N]$\sim 0$. After FDU, [C/N] decreases by the amount 
fixed by its current mass -- like a
single star --, thus moving downwards to the blue sequence of panel B in Fig.~\ref{cn_ratio}, 
equivalent to the lower sequence seen in Fig.~\ref{cn_stats}.  

Lithium is another element that depletes after FDU.  It is thus possible that some stars with [C/N]$\sim 0$ (pre-FDU) are also Li-normal.  The chemical abundance analysis of  \cite{2015A&A...584L...3J}  showed that the star $Y1$ is Li-enhanced to normal ($\mathrm{A(Li)} = 1.8$). If we assume (i) the abundance of Li of $Y1$ is pristine of a {pre-}FDU star, and (ii) $Y1$ is not a binary star, then $Y1$ could be one of the truly young $\alpha-$rich stars proposed by \cite{martig15} and \cite{2015A&A...576L..12C}.  $Y1$ could as well be a merger product as those suggested by \cite{2015ApJ...807...82T}, although in this case the star should have lost angular momentum as its rotational velocity is low \citep[$v \sin i \sim 1~\mathrm{km/s}$, ][]{2015A&A...584L...3J}.  It has been however suggested that mass transfer in blue stragglers should rather destroy Li \citep{2005AJ....129..466C}, although the picture is not so simple since other mass-transfer products show a large scatter in Li, from Li-depleted to Li-rich \citep[see e.g. discussions in][Hansen et al, {\it subm.}]{2012ApJ...751...14M}.   

\section{Conclusions} 

In this article we have shown that binary stars { may also} explain the nature of the ``young" $\alpha-$rich stars.  Our observations reveal that the sample used by \cite{martig15} to show that these stars are young contains { a substantial percentage of binaries (at least 50\%)}, and therefore isochrones of individual objects should not be used for determining the age of these stars.  In addition, our population synthesis model shows that it is possible to obtain the (high) masses and [C/N] abundances of these stars by  mass-transfer in binary systems. A more detailed population synthesis study focusing in a more quantitative way on the binary fraction and mass distribution of the \apo\ sample is on-going (Izzard et al., in preparation). Furthermore,  a { long-term} monitoring of the RVs of our sample is still on-going, with the goal to obtain accurate orbital properties even for long-period binaries. { Parameters like the orbital eccentricity, and more generally the location of the systems in the eccentricity -- period diagram, provide crucial help in understanding the process of mass transfer.} RV monitoring takes time \citep{2009PASA...26..372P}, thus patience is the key ingredient needed to disentangle the origin of these interesting objects and finally quantify how many of them are truly young stars.

 As  \cite{1979ApJ...234..569W} wrote, BSSs  seem to be both inadequately understood and insufficiently appreciated.  However, we know that as long as the possibility of mass transfer exists, blue stragglers will form. Because many show no evident signatures in their spectra { \citep{2016MNRAS.tmp..463Y} } and because they might be the product of  coalescence resulting in single objects, identifying those BSSs that have evolved off the main-sequence has been so far an almost impossible mission.  Spectro-seismic surveys give us a unique opportunity to study them,  opening a new window to the physics of stellar evolution and mass transfer. \\
%

\section*{Acknowledgements}
This work was partly supported by the European Union FP7 programme through ERC grant number 320360.  PJ acknowledges King's College Cambridge for partially supporting this work.  We thank C. Tout and S. Aarseth for discussion on the subject. KH is supported by Marshall Scholarship and King's College Cambridge Studenship. RJI thanks the STFC for funding his Rutherford Fellowship. Based on observations made with the Mercator Telescope, operated on the island of La Palma by the Flemish Community, at the Spanish Observatorio del Roque de los Muchachos of the Instituto de Astrof\' i sica de Canarias. Based on observations obtained with the HERMES spectrograph, which is supported by the Research Foundation - Flanders (FWO), Belgium, the Research Council of KU Leuven, Belgium, the Fonds National de la Recherche Scientifique (F.R.S.-FNRS), Belgium, the Royal Observatory of Belgium, the Observatoire de Gen\`{e}ve, Switzerland and the Th\"{u}ringer Landessternwarte Tautenburg, Germany.

\bibliographystyle{aa} 
\bibliography{refs_youngalpha} 

\begin{thebibliography}{68}
\expandafter\ifx\csname natexlab\endcsname\relax\def\natexlab#1{#1}\fi

\bibitem[{{Abate} {et~al.}(2013){Abate}, {Pols}, {Izzard}, {Mohamed}, \& {de
  Mink}}]{2013A&A...552A..26A}
{Abate}, C., {Pols}, O.~R., {Izzard}, R.~G., {Mohamed}, S.~S., \& {de Mink},
  S.~E. 2013, \aap, 552, A26

\bibitem[{{Abate} {et~al.}(2015){Abate}, {Pols}, {Stancliffe}, {Izzard},
  {Karakas}, {Beers}, \& {Lee}}]{2015A&A...581A..62A}
{Abate}, C., {Pols}, O.~R., {Stancliffe}, R.~J., {et~al.} 2015, \aap, 581, A62

\bibitem[{{Anders} {et~al.}(2014){Anders}, {Chiappini}, {Santiago},
  {Rocha-Pinto}, {Girardi}, {da Costa}, {Maia}, {Steinmetz}, {Minchev},
  {Schultheis}, {Boeche}, {Miglio}, {Montalb{\'a}n}, {Schneider}, {Beers},
  {Cunha}, {Allende Prieto}, {Balbinot}, {Bizyaev}, {Brauer}, {Brinkmann},
  {Frinchaboy}, {Garc{\'{\i}}a P{\'e}rez}, {Hayden}, {Hearty}, {Holtzman},
  {Johnson}, {Kinemuchi}, {Majewski}, {Malanushenko}, {Malanushenko},
  {Nidever}, {O'Connell}, {Pan}, {Robin}, {Schiavon}, {Shetrone}, {Skrutskie},
  {Smith}, {Stassun}, \& {Zasowski}}]{2014A&A...564A.115A}
{Anders}, F., {Chiappini}, C., {Santiago}, B.~X., {et~al.} 2014, \aap, 564,
  A115

\bibitem[{{Baglin} {et~al.}(2006){Baglin}, {Auvergne}, {Boisnard}, {Lam-Trong},
  {Barge}, {Catala}, {Deleuil}, {Michel}, \& {Weiss}}]{2006cosp...36.3749B}
{Baglin}, A., {Auvergne}, M., {Boisnard}, L., {et~al.} 2006, in COSPAR Meeting,
  Vol.~36, 36th COSPAR Scientific Assembly

\bibitem[{{Bidelman} \& {Keenan}(1951)}]{1951ApJ...114..473B}
{Bidelman}, W.~P. \& {Keenan}, P.~C. 1951, \apj, 114, 473

\bibitem[{{Boffin} {et~al.}(2015){Boffin}, {Carraro}, \&
  {Beccari}}]{2015ebss.book.....B}
{Boffin}, H.~M.~J., {Carraro}, G., \& {Beccari}, G. 2015, {Ecology of Blue
  Straggler Stars}

\bibitem[{{Borucki} {et~al.}(2010){Borucki}, {Koch}, {Basri}, {Batalha},
  {Brown}, {Caldwell}, {Caldwell}, {Christensen-Dalsgaard}, {Cochran},
  {DeVore}, {Dunham}, {Dupree}, {Gautier}, {Geary}, {Gilliland}, {Gould},
  {Howell}, {Jenkins}, {Kondo}, {Latham}, {Marcy}, {Meibom}, {Kjeldsen},
  {Lissauer}, {Monet}, {Morrison}, {Sasselov}, {Tarter}, {Boss}, {Brownlee},
  {Owen}, {Buzasi}, {Charbonneau}, {Doyle}, {Fortney}, {Ford}, {Holman},
  {Seager}, {Steffen}, {Welsh}, {Rowe}, {Anderson}, {Buchhave}, {Ciardi},
  {Walkowicz}, {Sherry}, {Horch}, {Isaacson}, {Everett}, {Fischer}, {Torres},
  {Johnson}, {Endl}, {MacQueen}, {Bryson}, {Dotson}, {Haas}, {Kolodziejczak},
  {Van Cleve}, {Chandrasekaran}, {Twicken}, {Quintana}, {Clarke}, {Allen},
  {Li}, {Wu}, {Tenenbaum}, {Verner}, {Bruhweiler}, {Barnes}, \&
  {Prsa}}]{2010Sci...327..977B}
{Borucki}, W.~J., {Koch}, D., {Basri}, G., {et~al.} 2010, Science, 327, 977

\bibitem[{{Brogaard} {et~al.}(2016){Brogaard}, {Jessen-Hansen}, {Handberg},
  {Arentoft}, {Frandsen}, {Grundahl}, {Bruntt}, {Sandquist}, {Miglio}, {Beck},
  {Thygesen}, {Kj{\ae}rgaard}, \& {Haugaard}}]{2016arXiv160101412B}
{Brogaard}, K., {Jessen-Hansen}, J., {Handberg}, R., {et~al.} 2016, ArXiv
  e-prints

\bibitem[{{Cantat-Gaudin} {et~al.}(2014){Cantat-Gaudin}, {Vallenari}, {Zaggia},
  {Bragaglia}, {Sordo}, {Drew}, {Eisloeffel}, {Farnhill}, {Gonzalez-Solares},
  {Greimel}, {Irwin}, {Kupcu-Yoldas}, {Jordi}, {Blomme}, {Sampedro}, {Costado},
  {Alfaro}, {Smiljanic}, {Magrini}, {Donati}, {Friel}, {Jacobson}, {Abbas},
  {Hatzidimitriou}, {Spagna}, {Vecchiato}, {Balaguer-Nunez}, {Lardo}, {Tosi},
  {Pancino}, {Klutsch}, {Tautvaisiene}, {Drazdauskas}, {Puzeras},
  {Jim{\'e}nez-Esteban}, {Maiorca}, {Geisler}, {San Roman}, {Villanova},
  {Gilmore}, {Randich}, {Bensby}, {Flaccomio}, {Lanzafame}, {Recio-Blanco},
  {Damiani}, {Hourihane}, {Jofr{\'e}}, {de Laverny}, {Masseron}, {Morbidelli},
  {Prisinzano}, {Sacco}, {Sbordone}, \& {Worley}}]{2014A&A...569A..17C}
{Cantat-Gaudin}, T., {Vallenari}, A., {Zaggia}, S., {et~al.} 2014, \aap, 569,
  A17

\bibitem[{{Carney} {et~al.}(2005){Carney}, {Latham}, \&
  {Laird}}]{2005AJ....129..466C}
{Carney}, B.~W., {Latham}, D.~W., \& {Laird}, J.~B. 2005, \aj, 129, 466

\bibitem[{{Carney} {et~al.}(2001){Carney}, {Latham}, {Laird}, {Grant}, \&
  {Morse}}]{2001AJ....122.3419C}
{Carney}, B.~W., {Latham}, D.~W., {Laird}, J.~B., {Grant}, C.~E., \& {Morse},
  J.~A. 2001, \aj, 122, 3419

\bibitem[{{Carney} {et~al.}(2008){Carney}, {Latham}, {Stefanik}, \&
  {Laird}}]{2008AJ....135..196C}
{Carney}, B.~W., {Latham}, D.~W., {Stefanik}, R.~P., \& {Laird}, J.~B. 2008,
  \aj, 135, 196

\bibitem[{{Chaplin} {et~al.}(2011){Chaplin}, {Kjeldsen},
  {Christensen-Dalsgaard}, {Basu}, {Miglio}, {Appourchaux}, {Bedding},
  {Elsworth}, {Garc{\'{\i}}a}, {Gilliland}, {Girardi}, {Houdek}, {Karoff},
  {Kawaler}, {Metcalfe}, {Molenda-{\.Z}akowicz}, {Monteiro}, {Thompson},
  {Verner}, {Ballot}, {Bonanno}, {Brand{\~a}o}, {Broomhall}, {Bruntt},
  {Campante}, {Corsaro}, {Creevey}, {Do{\u g}an}, {Esch}, {Gai}, {Gaulme},
  {Hale}, {Handberg}, {Hekker}, {Huber}, {Jim{\'e}nez}, {Mathur}, {Mazumdar},
  {Mosser}, {New}, {Pinsonneault}, {Pricopi}, {Quirion}, {R{\'e}gulo},
  {Salabert}, {Serenelli}, {Silva Aguirre}, {Sousa}, {Stello}, {Stevens},
  {Suran}, {Uytterhoeven}, {White}, {Borucki}, {Brown}, {Jenkins}, {Kinemuchi},
  {Van Cleve}, \& {Klaus}}]{2011Sci...332..213C}
{Chaplin}, W.~J., {Kjeldsen}, H., {Christensen-Dalsgaard}, J., {et~al.} 2011,
  Science, 332, 213

\bibitem[{{Chen} \& {Han}(2008{\natexlab{a}})}]{2008MNRAS.384.1263C}
{Chen}, X. \& {Han}, Z. 2008{\natexlab{a}}, \mnras, 384, 1263

\bibitem[{{Chen} \& {Han}(2008{\natexlab{b}})}]{2008MNRAS.387.1416C}
{Chen}, X. \& {Han}, Z. 2008{\natexlab{b}}, \mnras, 387, 1416

\bibitem[{{Chiappini} {et~al.}(2015){Chiappini}, {Anders}, {Rodrigues},
  {Miglio}, {Montalb{\'a}n}, {Mosser}, {Girardi}, {Valentini}, {Noels},
  {Morel}, {Minchev}, {Steinmetz}, {Santiago}, {Schultheis}, {Martig}, {da
  Costa}, {Maia}, {Allende Prieto}, {de Assis Peralta}, {Hekker},
  {Theme{\ss}l}, {Kallinger}, {Garc{\'{\i}}a}, {Mathur}, {Baudin}, {Beers},
  {Cunha}, {Harding}, {Holtzman}, {Majewski}, {M{\'e}sz{\'a}ros}, {Nidever},
  {Pan}, {Schiavon}, {Shetrone}, {Schneider}, \&
  {Stassun}}]{2015A&A...576L..12C}
{Chiappini}, C., {Anders}, F., {Rodrigues}, T.~S., {et~al.} 2015, \aap, 576,
  L12

\bibitem[{{Chubak} {et~al.}(2012){Chubak}, {Marcy}, {Fischer}, {Howard},
  {Isaacson}, {Johnson}, \& {Wright}}]{2012arXiv1207.6212C}
{Chubak}, C., {Marcy}, G., {Fischer}, D.~A., {et~al.} 2012, ArXiv e-prints

\bibitem[{{Claeys} {et~al.}(2014){Claeys}, {Pols}, {Izzard}, {Vink}, \&
  {Verbunt}}]{2014A&A...563A..83C}
{Claeys}, J.~S.~W., {Pols}, O.~R., {Izzard}, R.~G., {Vink}, J., \& {Verbunt},
  F.~W.~M. 2014, \aap, 563, A83

\bibitem[{{Duquennoy} \& {Mayor}(1991)}]{1991A&A...248..485D}
{Duquennoy}, A. \& {Mayor}, M. 1991, \aap, 248, 485

\bibitem[{{Eisenstein} {et~al.}(2011){Eisenstein}, {Weinberg}, {Agol},
  {Aihara}, {Allende Prieto}, {Anderson}, {Arns}, {Aubourg}, {Bailey},
  {Balbinot}, \& et~al.}]{2011AJ....142...72E}
{Eisenstein}, D.~J., {Weinberg}, D.~H., {Agol}, E., {et~al.} 2011, \aj, 142, 72

\bibitem[{{Famaey} {et~al.}(2009){Famaey}, {Pourbaix}, {Frankowski}, {van Eck},
  {Mayor}, {Udry}, \& {Jorissen}}]{2009A&A...498..627F}
{Famaey}, B., {Pourbaix}, D., {Frankowski}, A., {et~al.} 2009, \aap, 498, 627

\bibitem[{{Frankowski} {et~al.}(2009){Frankowski}, {Famaey}, {van Eck},
  {Mayor}, {Udry}, \& {Jorissen}}]{2009A&A...498..479F}
{Frankowski}, A., {Famaey}, B., {van Eck}, S., {et~al.} 2009, \aap, 498, 479

\bibitem[{{Geller} \& {Mathieu}(2012)}]{2012AJ....144...54G}
{Geller}, A.~M. \& {Mathieu}, R.~D. 2012, \aj, 144, 54

\bibitem[{{Green}(2011)}]{2011BASI...39..289G}
{Green}, D.~A. 2011, Bulletin of the Astronomical Society of India, 39, 289

\bibitem[{{Hawkins} {et~al.}(2016){Hawkins}, {Masseron}, {Jofre}, {Gilmore},
  {Elsworth}, \& {Hekker}}]{2016arXiv160408800H}
{Hawkins}, K., {Masseron}, T., {Jofre}, P., {et~al.} 2016, ArXiv e-prints

\bibitem[{{Haywood} {et~al.}(2013){Haywood}, {Di Matteo}, {Lehnert}, {Katz}, \&
  {G{\'o}mez}}]{2013A&A...560A.109H}
{Haywood}, M., {Di Matteo}, P., {Lehnert}, M.~D., {Katz}, D., \& {G{\'o}mez},
  A. 2013, \aap, 560, A109

\bibitem[{{Holtzman} {et~al.}(2015){Holtzman}, {Shetrone}, {Johnson}, {Allende
  Prieto}, {Anders}, {Andrews}, {Beers}, {Bizyaev}, {Blanton}, {Bovy},
  {Carrera}, {Chojnowski}, {Cunha}, {Eisenstein}, {Feuillet}, {Frinchaboy},
  {Galbraith-Frew}, {Garc{\'{\i}}a P{\'e}rez}, {Garc{\'{\i}}a-Hern{\'a}ndez},
  {Hasselquist}, {Hayden}, {Hearty}, {Ivans}, {Majewski}, {Martell},
  {Meszaros}, {Muna}, {Nidever}, {Nguyen}, {O'Connell}, {Pan}, {Pinsonneault},
  {Robin}, {Schiavon}, {Shane}, {Sobeck}, {Smith}, {Troup}, {Weinberg},
  {Wilson}, {Wood-Vasey}, {Zamora}, \& {Zasowski}}]{2015AJ....150..148H}
{Holtzman}, J.~A., {Shetrone}, M., {Johnson}, J.~A., {et~al.} 2015, \aj, 150,
  148

\bibitem[{{Iben}(1965)}]{1965ApJ...142.1447I}
{Iben}, Jr., I. 1965, \apj, 142, 1447

\bibitem[{{Izzard} {et~al.}(2006){Izzard}, {Dray}, {Karakas}, {Lugaro}, \&
  {Tout}}]{2006A&A...460..565I}
{Izzard}, R.~G., {Dray}, L.~M., {Karakas}, A.~I., {Lugaro}, M., \& {Tout},
  C.~A. 2006, \aap, 460, 565

\bibitem[{{Izzard} {et~al.}(2009){Izzard}, {Glebbeek}, {Stancliffe}, \&
  {Pols}}]{2009A&A...508.1359I}
{Izzard}, R.~G., {Glebbeek}, E., {Stancliffe}, R.~J., \& {Pols}, O.~R. 2009,
  \aap, 508, 1359

\bibitem[{{Izzard} {et~al.}(2004){Izzard}, {Tout}, {Karakas}, \&
  {Pols}}]{2004MNRAS.350..407I}
{Izzard}, R.~G., {Tout}, C.~A., {Karakas}, A.~I., \& {Pols}, O.~R. 2004,
  \mnras, 350, 407

\bibitem[{{Jofr{\'e}} {et~al.}(2015{\natexlab{a}}){Jofr{\'e}}, {Petrucci},
  {Garc{\'{\i}}a}, \& {G{\'o}mez}}]{2015A&A...584L...3J}
{Jofr{\'e}}, E., {Petrucci}, R., {Garc{\'{\i}}a}, L., \& {G{\'o}mez}, M.
  2015{\natexlab{a}}, \aap, 584, L3

\bibitem[{{Jofr{\'e}} {et~al.}(2015{\natexlab{b}}){Jofr{\'e}}, {Heiter},
  {Soubiran}, {Blanco-Cuaresma}, {Masseron}, {Nordlander}, {Chemin}, {Worley},
  {Van Eck}, {Hourihane}, {Gilmore}, {Adibekyan}, {Bergemann}, {Cantat-Gaudin},
  {Delgado-Mena}, {Gonz{\'a}lez Hern{\'a}ndez}, {Guiglion}, {Lardo}, {de
  Laverny}, {Lind}, {Magrini}, {Mikolaitis}, {Montes}, {Pancino},
  {Recio-Blanco}, {Sordo}, {Sousa}, {Tabernero}, \&
  {Vallenari}}]{2015A&A...582A..81J}
{Jofr{\'e}}, P., {Heiter}, U., {Soubiran}, C., {et~al.} 2015{\natexlab{b}},
  \aap, 582, A81

\bibitem[{{Jofr{\'e}} {et~al.}(2014){Jofr{\'e}}, {Heiter}, {Soubiran},
  {Blanco-Cuaresma}, {Worley}, {Pancino}, {Cantat-Gaudin}, {Magrini},
  {Bergemann}, {Gonz{\'a}lez Hern{\'a}ndez}, {Hill}, {Lardo}, {de Laverny},
  {Lind}, {Masseron}, {Montes}, {Mucciarelli}, {Nordlander}, {Recio Blanco},
  {Sobeck}, {Sordo}, {Sousa}, {Tabernero}, {Vallenari}, \& {Van
  Eck}}]{2014A&A...564A.133J}
{Jofr{\'e}}, P., {Heiter}, U., {Soubiran}, C., {et~al.} 2014, \aap, 564, A133

\bibitem[{{Jofr{\'e}} \& {Weiss}(2011)}]{2011A&A...533A..59J}
{Jofr{\'e}}, P. \& {Weiss}, A. 2011, \aap, 533, A59

\bibitem[{{Jorissen} \& {Mayor}(1988)}]{1988A&A...198..187J}
{Jorissen}, A. \& {Mayor}, M. 1988, \aap, 198, 187

\bibitem[{{Jorissen} {et~al.}(1998){Jorissen}, {Van Eck}, {Mayor}, \&
  {Udry}}]{1998A&A...332..877J}
{Jorissen}, A., {Van Eck}, S., {Mayor}, M., \& {Udry}, S. 1998, \aap, 332, 877

\bibitem[{{Jorissen} {et~al.}(2016){Jorissen}, {Van Eck}, {Van Winckel},
  {Merle}, {Boffin}, {Andersen}, {Nordstr{\"o}m}, {Udry}, {Masseron},
  {Lenaerts}, \& {Waelkens}}]{2016A&A...586A.158J}
{Jorissen}, A., {Van Eck}, S., {Van Winckel}, H., {et~al.} 2016, \aap, 586,
  A158

\bibitem[{{Keenan}(1942)}]{1942ApJ....96..101K}
{Keenan}, P.~C. 1942, \apj, 96, 101

\bibitem[{{Lagarde} {et~al.}(2012){Lagarde}, {Decressin}, {Charbonnel},
  {Eggenberger}, {Ekstr{\"o}m}, \& {Palacios}}]{2012A&A...543A.108L}
{Lagarde}, N., {Decressin}, T., {Charbonnel}, C., {et~al.} 2012, \aap, 543,
  A108

\bibitem[{{Lucatello} {et~al.}(2005){Lucatello}, {Tsangarides}, {Beers},
  {Carretta}, {Gratton}, \& {Ryan}}]{2005ApJ...625..825L}
{Lucatello}, S., {Tsangarides}, S., {Beers}, T.~C., {et~al.} 2005, \apj, 625,
  825

\bibitem[{{Magrini} {et~al.}(2015){Magrini}, {Randich}, {Donati}, {Bragaglia},
  {Adibekyan}, {Romano}, {Smiljanic}, {Blanco-Cuaresma}, {Tautvai{\v
  s}ien{\.e}}, {Friel}, {Overbeek}, {Jacobson}, {Cantat-Gaudin}, {Vallenari},
  {Sordo}, {Pancino}, {Geisler}, {San Roman}, {Villanova}, {Casey},
  {Hourihane}, {Worley}, {Francois}, {Gilmore}, {Bensby}, {Flaccomio}, {Korn},
  {Recio-Blanco}, {Carraro}, {Costado}, {Franciosini}, {Heiter}, {Jofr{\'e}},
  {Lardo}, {de Laverny}, {Monaco}, {Morbidelli}, {Sacco}, {Sousa}, \&
  {Zaggia}}]{2015A&A...580A..85M}
{Magrini}, L., {Randich}, S., {Donati}, P., {et~al.} 2015, \aap, 580, A85

\bibitem[{{Martig} {et~al.}(2016){Martig}, {Fouesneau}, {Rix}, {Ness},
  {M{\'e}sz{\'a}ros}, {Garc{\'{\i}}a-Hern{\'a}ndez}, {Pinsonneault},
  {Serenelli}, {Aguirre}, \& {Zamora}}]{2016MNRAS.456.3655M}
{Martig}, M., {Fouesneau}, M., {Rix}, H.-W., {et~al.} 2016, \mnras, 456, 3655

\bibitem[{{Martig} {et~al.}(2015){Martig}, {Rix}, {Aguirre}, {Hekker},
  {Mosser}, {Elsworth}, {Bovy}, {Stello}, {Anders}, {Garc{\'{\i}}a}, {Tayar},
  {Rodrigues}, {Basu}, {Carrera}, {Ceillier}, {Chaplin}, {Chiappini},
  {Frinchaboy}, {Garc{\'{\i}}a-Hern{\'a}ndez}, {Hearty}, {Holtzman}, {Johnson},
  {Majewski}, {Mathur}, {M{\'e}sz{\'a}ros}, {Miglio}, {Nidever}, {Pan},
  {Pinsonneault}, {Schiavon}, {Schneider}, {Serenelli}, {Shetrone}, \&
  {Zamora}}]{martig15}
{Martig}, M., {Rix}, H.-W., {Aguirre}, V.~S., {et~al.} 2015, \mnras, 451, 2230

\bibitem[{{Masseron} \& {Gilmore}(2015)}]{2015MNRAS.453.1855M}
{Masseron}, T. \& {Gilmore}, G. 2015, \mnras, 453, 1855

\bibitem[{{Masseron} {et~al.}(2012){Masseron}, {Johnson}, {Lucatello},
  {Karakas}, {Plez}, {Beers}, \& {Christlieb}}]{2012ApJ...751...14M}
{Masseron}, T., {Johnson}, J.~A., {Lucatello}, S., {et~al.} 2012, \apj, 751, 14

\bibitem[{{McClure} \& {Woodsworth}(1990)}]{1990ApJ...352..709M}
{McClure}, R.~D. \& {Woodsworth}, A.~W. 1990, \apj, 352, 709

\bibitem[{{McCrea}(1964)}]{1964MNRAS.128..147M}
{McCrea}, W.~H. 1964, \mnras, 128, 147

\bibitem[{{Minchev} {et~al.}(2013){Minchev}, {Chiappini}, \&
  {Martig}}]{2013A&A...558A...9M}
{Minchev}, I., {Chiappini}, C., \& {Martig}, M. 2013, \aap, 558, A9

\bibitem[{{Minchev} {et~al.}(2014){Minchev}, {Chiappini}, {Martig},
  {Steinmetz}, {de Jong}, {Boeche}, {Scannapieco}, {Zwitter}, {Wyse}, {Binney},
  {Bland-Hawthorn}, {Bienaym{\'e}}, {Famaey}, {Freeman}, {Gibson}, {Grebel},
  {Gilmore}, {Helmi}, {Kordopatis}, {Lee}, {Munari}, {Navarro}, {Parker},
  {Quillen}, {Reid}, {Siebert}, {Siviero}, {Seabroke}, {Watson}, \&
  {Williams}}]{2014ApJ...781L..20M}
{Minchev}, I., {Chiappini}, C., {Martig}, M., {et~al.} 2014, \apjl, 781, L20

\bibitem[{{Nidever} {et~al.}(2015){Nidever}, {Holtzman}, {Allende Prieto},
  {Beland}, {Bender}, {Bizyaev}, {Burton}, {Desphande}, {Fleming},
  {Garc{\'{\i}}a P{\'e}rez}, {Hearty}, {Majewski}, {M{\'e}sz{\'a}ros}, {Muna},
  {Nguyen}, {Schiavon}, {Shetrone}, {Skrutskie}, {Sobeck}, \&
  {Wilson}}]{2015AJ....150..173N}
{Nidever}, D.~L., {Holtzman}, J.~A., {Allende Prieto}, C., {et~al.} 2015, \aj,
  150, 173

\bibitem[{{Pinsonneault} {et~al.}(2014){Pinsonneault}, {Elsworth}, {Epstein},
  {Hekker}, {M{\'e}sz{\'a}ros}, {Chaplin}, {Johnson}, {Garc{\'{\i}}a},
  {Holtzman}, {Mathur}, {Garc{\'{\i}}a P{\'e}rez}, {Silva Aguirre}, {Girardi},
  {Basu}, {Shetrone}, {Stello}, {Allende Prieto}, {An}, {Beck}, {Beers},
  {Bizyaev}, {Bloemen}, {Bovy}, {Cunha}, {De Ridder}, {Frinchaboy},
  {Garc{\'{\i}}a-Hern{\'a}ndez}, {Gilliland}, {Harding}, {Hearty}, {Huber},
  {Ivans}, {Kallinger}, {Majewski}, {Metcalfe}, {Miglio}, {Mosser}, {Muna},
  {Nidever}, {Schneider}, {Serenelli}, {Smith}, {Tayar}, {Zamora}, \&
  {Zasowski}}]{2014ApJS..215...19P}
{Pinsonneault}, M.~H., {Elsworth}, Y., {Epstein}, C., {et~al.} 2014, \apjs,
  215, 19

\bibitem[{{Piotto} {et~al.}(2004){Piotto}, {De Angeli}, {King}, {Djorgovski},
  {Bono}, {Cassisi}, {Meylan}, {Recio-Blanco}, {Rich}, \&
  {Davies}}]{2004ApJ...604L.109P}
{Piotto}, G., {De Angeli}, F., {King}, I.~R., {et~al.} 2004, \apjl, 604, L109

\bibitem[{{Preston}(2009)}]{2009PASA...26..372P}
{Preston}, G.~W. 2009, \pasa, 26, 372

\bibitem[{{Preston}(2015)}]{2015ebss.book...65P}
{Preston}, G.~W. 2015, {Field Blue Stragglers and Related Mass Transfer
  Issues}, ed. H.~M.~J. {Boffin}, G.~{Carraro}, \& G.~{Beccari}, 65

\bibitem[{{Preston} \& {Sneden}(2000)}]{2000AJ....120.1014P}
{Preston}, G.~W. \& {Sneden}, C. 2000, \aj, 120, 1014

\bibitem[{{Raskin} {et~al.}(2011){Raskin}, {van Winckel}, {Hensberge},
  {Jorissen}, {Lehmann}, {Waelkens}, {Avila}, {de Cuyper}, {Degroote},
  {Dubosson}, {Dumortier}, {Fr{\'e}mat}, {Laux}, {Michaud}, {Morren}, {Perez
  Padilla}, {Pessemier}, {Prins}, {Smolders}, {van Eck}, \&
  {Winkler}}]{2011A&A...526A..69R}
{Raskin}, G., {van Winckel}, H., {Hensberge}, H., {et~al.} 2011, \aap, 526, A69

\bibitem[{{Salaris} {et~al.}(2015){Salaris}, {Pietrinferni}, {Piersimoni}, \&
  {Cassisi}}]{2015A&A...583A..87S}
{Salaris}, M., {Pietrinferni}, A., {Piersimoni}, A.~M., \& {Cassisi}, S. 2015,
  \aap, 583, A87

\bibitem[{{Sandage}(1953)}]{1953AJ.....58...61S}
{Sandage}, A.~R. 1953, \aj, 58, 61

\bibitem[{{Santucci} {et~al.}(2015){Santucci}, {Placco}, {Rossi}, {Beers},
  {Reggiani}, {Lee}, {Xue}, \& {Carollo}}]{2015ApJ...801..116S}
{Santucci}, R.~M., {Placco}, V.~M., {Rossi}, S., {et~al.} 2015, \apj, 801, 116

\bibitem[{{Setiawan} {et~al.}(2003){Setiawan}, {Pasquini}, {da Silva}, {von der
  L{\"u}he}, \& {Hatzes}}]{2003A&A...397.1151S}
{Setiawan}, J., {Pasquini}, L., {da Silva}, L., {von der L{\"u}he}, O., \&
  {Hatzes}, A. 2003, \aap, 397, 1151

\bibitem[{{Tautvai{\v s}ien{\.e}} {et~al.}(2015){Tautvai{\v s}ien{\.e}},
  {Drazdauskas}, {Mikolaitis}, {Barisevi{\v c}ius}, {Puzeras}, {Stonkut{\.e}},
  {Chorniy}, {Magrini}, {Romano}, {Smiljanic}, {Bragaglia}, {Carraro}, {Friel},
  {Morel}, {Pancino}, {Donati}, {Jim{\'e}nez-Esteban}, {Gilmore}, {Randich},
  {Jeffries}, {Vallenari}, {Bensby}, {Flaccomio}, {Recio-Blanco}, {Costado},
  {Hill}, {Jofr{\'e}}, {Lardo}, {de Laverny}, {Masseron}, {Moribelli}, {Sousa},
  \& {Zaggia}}]{2015A&A...573A..55T}
{Tautvai{\v s}ien{\.e}}, G., {Drazdauskas}, A., {Mikolaitis}, {\v S}., {et~al.}
  2015, \aap, 573, A55

\bibitem[{{Tayar} {et~al.}(2015){Tayar}, {Ceillier},
  {Garc{\'{\i}}a-Hern{\'a}ndez}, {Troup}, {Mathur}, {Garc{\'{\i}}a}, {Zamora},
  {Johnson}, {Pinsonneault}, {M{\'e}sz{\'a}ros}, {Allende Prieto}, {Chaplin},
  {Elsworth}, {Hekker}, {Nidever}, {Salabert}, {Schneider}, {Serenelli},
  {Shetrone}, \& {Stello}}]{2015ApJ...807...82T}
{Tayar}, J., {Ceillier}, T., {Garc{\'{\i}}a-Hern{\'a}ndez}, D.~A., {et~al.}
  2015, \apj, 807, 82

\bibitem[{{Udry} {et~al.}(1999){Udry}, {Mayor}, \&
  {Queloz}}]{1999ASPC..185..367U}
{Udry}, S., {Mayor}, M., \& {Queloz}, D. 1999, in Astronomical Society of the
  Pacific Conference Series, Vol. 185, IAU Colloq. 170: Precise Stellar Radial
  Velocities, ed. J.~B. {Hearnshaw} \& C.~D. {Scarfe}, 367

\bibitem[{{Unavane} {et~al.}(1996){Unavane}, {Wyse}, \&
  {Gilmore}}]{1996MNRAS.278..727U}
{Unavane}, M., {Wyse}, R.~F.~G., \& {Gilmore}, G. 1996, \mnras, 278, 727

\bibitem[{{Wheeler}(1979)}]{1979ApJ...234..569W}
{Wheeler}, J.~C. 1979, \apj, 234, 569

\bibitem[{{Wilson} \& {Hilferty}(1931)}]{Wilson1931}
{Wilson}, E.~B. \& {Hilferty}, M.~M. 1931, Proc. Natl. Acad. Sci., 17, 684

\bibitem[{{Yong} {et~al.}(2016){Yong}, {Casagrande}, {Venn}, {Chen{\'e}},
  {Keown}, {Malo}, {Martioli}, {Alves-Brito}, {Asplund}, {Dotter}, {Martell},
  {Mel{\'e}ndez}, \& {Schlesinger}}]{2016MNRAS.tmp..463Y}
{Yong}, D., {Casagrande}, L., {Venn}, K.~A., {et~al.} 2016, \mnras

\end{thebibliography}
\onecolumn

\begin{appendix}
\section{Individual radial velocity measurements}
{In this appendix we list the radial velocities we have used for the analysis presented in this work. The values obtained in 2011 and 2012 are taken from the individual visits APOGEE Survey and the RVs have been determined by  \cite{2015AJ....150..173N}. The values obtained in 2015 and 2016  are taken from HERMES observations, and have been determined by us. Table~\ref{RV_individual} has two main columns, the left-hand column shows the RVs of the $Y$ stars while the right-hand column shows the RVs for the $O$ stars. Note that every star has at least three independent observations, allowing to distinguish between an erroneous measurement and a possible binary.  }

\small

\begin{longtable}{c c c c | c c c c}
\caption{Radial velocity measurements for individual epochs.  The left hand side of the table lists the RV measurements for the young sample ($Y$), while the right hand side lists the measurements for the old sample ($O$).   }\label{RV_individual}\\
\hline \hline
star &  Obs. Date & RV & $\sigma$  & star & Obs. Date  & RV & $\sigma$ \\ 
\hline
\endfirsthead
\caption{continued.}\\
\hline\hline
star &  Obs. Date & RV & $\sigma$  & star & Obs. Date  & RV & $\sigma$ \\ 
\hline
\endhead
\hline
\endfoot

$Y1^a$ & 2011-11-11 &-5.506&0.0136  & $O1$ & 2011-10-10&8.420&0.0135\\
$Y1$ & 2015-07-29 & -32.334 & 0.006 &$O1$ &2015-07-29 & 0.710 & 0.012  \\
$Y1$ &2016-05-27 & -5.831 &0.011& $O1$ &2016-05-28 & 5.459& 0.005\\
$Y1$ & 2016-06-04 & -5.663 &0.013& $O1$ &2016-06-25 & 5.702& 0.011\\
$Y1$ &2016-06-04 & -5.894& 0.012 \\
$Y1$& 2016-06-06& -5.708 &0.010\\
$Y1$ & 2016-06-15 &-5.846 &0.009\\
$Y1$ & 2016-06-18 & -5.756& 0.03\\
$Y1$ & 2016-07-01 &  -5.814 & 0.006& $O1$ &2016-07-29 & 5.884& 0.013\\
$Y1$ & 2016-07-26 & -5.751  & 0.016& $O1$ &2016-08-01 & 6.001& 0.013\\ 
\hline
$Y2$ & 2011-10-07 & 6.494 &0.0134& $O2$ & 2011-10-07&-15.959&0.0148 \\
$Y2$ & 2011-11-06  & 6.563 & 0.0156& $O2$ & 2011-11-06&-15.930&0.0170 \\
$Y2$ & 2011-11-07  & 6.605 & 0.0112& $O2$ &2011-11-07&-15.991&0.0120 \\
$Y2$&2015-07-30 & 6.673&0.006 & $O2$ &2015-07-29 & -16.408 & 0.01 \\
$Y2$&2016-06-03 & 6.252 &0.007&$O2$& 2016-06-03 & -16.417& 0.01\\
$Y2$&2016-07-29 & 6.134 &0.008&$O2$& 2016-06-25 & -16.366& 0.012\\
         &                  &                   &  &$O2$& 2016-07-30 & -16.391& 0.007\\
         &                  &                   &  &$O2$& 2016-08-01 & -16.309& 0.013\\
\hline
$Y3$ & 2011-10-07 &-83.486&0.0117 & $O3$ & 2012-04-10&-22.169&0.0125 \\
$Y3$ & 2011-11-06 &-83.174&0.0151&   \\
$Y3$ & 2011-11-07 &-83.323&0.0106 &  \\
$Y3$ & 2015-07-30  & -83.415 & 0.008 & $O3$ &2015-07-25&-30.836 &0.005   \\
$Y3$ & 2015-08-02 & -83.384 & 0.006 & $O3$ & 2015-08-02& -30.426 &  0.008 \\
$Y3$ & 2016-05-29 &-83.288 &0.009 &$O3$ &2016-05-29 &-23.791& 0.004\\
$Y3$ & 2016-07-26 &-83.470 &0.014 & $O3$ &20160729 & -26.841&0.008\\
         &                  &                   &  & $O3$ &20160804 & -27.324&0.010\\
\hline
$Y4$ &  2011-10-10 & -75.510&0.0103& $O4$& 2011-09-19&-150.767&0.0348 \\
&&&& $O4$& 2011-10-06&-150.916&0.0424 \\
&&& &$O4$ &2011-10-17&-150.990&0.0352 \\
$Y4$ & 2015-07-30 & -69.434 & 0.008& $O4$  &  2015-07-29 & -151.150& 0.111\\
$Y4$ &2016-05-29 & -74.349& 0.006& \\
\hline
$Y5$ & 2012-05-12 & -85.965 & 0.0137 & $O5$ &2012-05-12&6.344&0.0088 \\
$Y5$ & 2012-05-15 & -85.997 & 0.0150 & $O5$ &2012-05-15&6.296&0.0095 \\
$Y5$ & 2012-05-26 & -85.894&0.0209 & $O5$ & 2012-05-26&6.361&0.0114 \\
$Y5$ & 2012-05-27 & -85.848&0.0175  & $O5$ & 2012-05-27&6.345&0.0096\\
$Y5$ & 2012-05-31 & -85.976&0.0154 & $O5$ & 2012-05-31&6.541&0.0140 \\
$Y5$ &2015-07-28& -86.246 &0.009 & $O5$ & 2015-07-29  & 5.986 & 0.005\\
$Y5$&2016-06-03 & -86.081& 0.014& &&&\\
$Y5$&2016-07-30 & -86.270& 0.006& $O5$ & 2016-07-29  & 6.044 & 0.006\\
\hline
$Y6$ &2012-05-25  & -18.209&0.0172& $O6$ & 2011-11-11&-30.587&0.0136 \\
$Y6$ &2012-05-30 & -17.911&0.0126 & & & & \\
$Y6$ & 2015-07-29&-19.005 &0.01& $O6$ &2015-07-28& -31.021 & 0.02\\
$Y6$ &2015-07-30 &-18.991 & 0.008& \\
$Y6$&2016-05-29 &-27.480& 0.011& $O6$ & 2016-05-29 &-30.848 &0.013\\
      &                   &        &          & $O6$ & 2016-06-25 & -30.958& 0.024\\
$Y6$&2016-07-29 &-27.998& 0.008&  $O6$  & 2016-07-30 & -30.822 & 0.019\\
$Y6$&2016-08-01 &-28.076& 0.011& $O6$ & 2016-08-01 & -31.005 & 0.019\\
\hline
$Y7$ & 2011-11-09&-39.063&0.0229 & $O7$ & 2012-05-25&2.318&0.0235 \\
&&&& $O7$ & 2012-05-30&2.415&0.0152 \\
$Y7$ &2015-07-29& -37.023& 0.01 & $O7$ &2015-07-29 & 1.934 & 0.014\\
$Y7$ &2015-07-29& -36.854&0.011 \\
$Y7$ & 2015-07-31& -36.825 & 0.024 \\
$Y7$ &  2016-06-06 & -39.444 & 0.058 & $O7$ &2016-05-29&2.013 & 0.015\\
$Y7$ & 2016-07-30& -40.755 & 0.025 & $O7$ &2016-07-29 &1.998 & 0.018\\
       &                   &        &          &$O7$ &2016-08-01 &1.937 & 0.013\\
\hline
$Y8$ &  2011-11-11&-39.532&0.0053& $O8$ &2012-04-08&-46.695&0.0113\\
&&&& $O8$ &2012-05-04&-46.489&0.0116\\
&&&& $O8$ &2013-06-09&-46.334&0.0105\\
$Y8$ & 2015-07-25& -42.747 & 0.006 & $O8$ &2015-07-25&-46.470 & 0.013 \\
$Y8$ & 2015-08-02 &-42.811& 0.008 & $O8$ & 2015-08-02& -46.569 & 0.019  \\
$Y8$&2016-05-29&-44.290 &0.008&$O8$ & 2016-05-29 &-46.462 &0.017\\
$Y8$&2016-06-18&-44.476 &0.021\\
$Y8$&2016-07-28&-44.745 &0.007&$O8$&2016-07-28&-46.547 &0.012\\
\hline
$Y9$ & 2012-05-12 & -37.575&0.0067 &$O9$ & 2011-09-07&-69.756&0.0053 \\
$Y9$ &2012-05-15  & -37.613&0.0073 & $O9$ & 2011-10-06&-69.798&0.0057 \\
$Y9$ &  2012-05-26& -37.736&0.0092& $O9$ & 2011-10-17&-69.760&0.0050\\
$Y9$ & 2012-05-27 &-37.664&0.0077 & & & & \\
$Y9$ &  2012-05-31&-37.655&0.0073&  & & & \\
$Y9$ & 2015-07-28& -37.828 & 0.004 & $O9$ &2015-07-29&-69.958& 0.01\\
$Y9$ & 2016-06-03 & -30.289 &0.01 & $O9$ & 2016-06-03&-69.773 &0.013 \\
$Y9$ & 2016-06-18 &  -30.222 &0.025 & $O9$ & 2016-06-25&-69.683 &0.010\\
$Y9$ & 2016-07-29 &  -30.490 &0.008& $O9$ & 2016-07-30&-69.994 &0.015 \\
       &                   &        &          &  $O9$ & 2016-08-01&-69.938 &0.015 \\
\hline
$Y10$ &  2011-11-11&-56.750&0.0058 & $O10$ &2011-09-09&6.762&0.0073 \\
&&&& $O10$ & 2011-10-15&6.719&0.0063 \\
&&&& $O10$ & 2011-11-01&6.725&0.0065\\
$Y10$ & 2015-07-25 & -57.027 & 0.007 & $O10$ & 2015-07-24& 6.863 & 0.012\\
$Y10$ & 2015-08-02 & -57.079 & 0.009 & $O10$ & 2015-08-02 & 6.949 &0.011\\
$Y10$ & 2016-06-06& -56.902 &0.007 & $O10$ & 2016-07-29 & 6.998 &0.008\\
$Y10$ & 2016-07-29& -56.970 &0.007\\
\hline
$Y11$ & 2011-10-10&-26.523&0.0095 &$O11$ & 2012-05-19&-5.750&0.0075 \\
&&&& $O11$ & 2012-05-20&-5.821&0.0165\\
&&&& $O11$ & 2012-05-21&-5.722&0.0077\\
$Y11$ &2015-07-29 & -26.759 & 0.009 & $O11$ & 2015-07-25&-5.993 &0.004 \\
$Y11$ &2015-08-01 & -26.791 & 0.009 & $O11$& 2015-08-02&-6.015&0.006 \\
$Y11$ & 2016-06-06&  -26.748 &0.017 & $O11$& 2015-07-28&-6.097&0.006\\
$Y11$ & 2016-07-29&  -26.752 &0.006 \\
\hline
$Y12$ &2011-09-19&-47.468&0.0051 & $O12$ & 2012-05-12&29.284&0.0133 \\
$Y12$ & 2011-10-06&-47.96&0.0060& $O12$ & 2012-05-15&29.332&0.0141 \\
$Y12$ &2011-10-17&-47.454&0.0048 & $O12$ & 2012-05-26&29.221&0.0175 \\
&&&& $O12$ &2012-05-27&29.327&0.0145\\
&&&& $O12$ &2012-05-31&29.386&0.0127\\
$Y12$ & 2015-07-29 &-46.559 & 0.006 & $O12$ & 2015-07-29 & 29.080 & 0.006 \\
$Y12$ & 2016-05-29 & -47.488 & 0.005 &$O12$& 2016-05-30 & 29.079 & 0.009 \\
$Y12$ & 2016-07-29 & -47.656 & 0.006 & $O12$ &2016-07-06&29.032&0.011\\
$Y12$ & 2016-07-30 & -47.616 & 0.006 & $O12$ &2016-07-29&29.267&0.020\\
$Y12$ & 2016-08-01 & -47.664 & 0.011 \\
\hline
$Y13$ &2011-10-07&-62.038&0.0028& $O13$ & 2012-05-19&-40.996&0.0092 \\
$Y13$ &  2011-11-06&-61.985&0.0034& $O13$ & 2012-05-20&-40.993&0.0197 \\
$Y13$ &2011-11-07&-61.951&0.0025 & $O13$ & 2012-05-21&-40.898&0.0097\\
$Y13$ &2015-07-29 & -62.277 & 0.003 & $O13$ &2015-07-28 & -41.531 & 0.007\\
         &                  &              &          & $O13$ & 2015-07-30&  -41.424 & 0.006\\
$Y13$& 2016-06-04&-62.201 &0.005& $O13$ & 2016-07-06 & -41.556 & 0.015\\
$Y13$& 2016-07-18&-62.079 &0.006& $O13$ & 2016-07-29&  -41.366 & 0.011\\
$Y13$& 2016-07-27&-62.123 &0.008\\
\medskip\\
\hline
\medskip\\


\noalign{\noindent Remark:
 a: Star $Y1$ is only  33'' westwards from star TYC 3542 434 1, which has a seemingly constant velocity of -32.35~km/s, measured 
on two occasions (2015-07-29 and 2016-07-25).}
\medskip\\
\end{longtable}

\end{appendix}




\end{document}